\documentclass[12pt]{article}

\usepackage{graphicx,amsmath,amsfonts,verbatim,amssymb,amsthm}

\theoremstyle{plain}

\newtheorem{corollary}{Corollary}

\theoremstyle{remark}
\newtheorem{remark}{Remark}
\newtheorem{example}{Example}

\let\ln=\log

\begin{document}

\title{
Probability Theory\\
Compatible with the New Conception\\
of Modern Thermodynamics.\\
Economics and Crisis of Debts}
\author{
V.~P.~Maslov\\
Steklov Institute of Mathematics, Russian Academy of Sciences,\\
117333 Moscow, Russia}

\date{November 3, 2011}

\maketitle

\begin{abstract}
We show that G\"odel's negative results concerning arithmetic, which date back to the
1930s, and the ancient ``sand pile'' paradox (known also as ``sorites paradox'') pose the questions
of the use of soft sets and of the effect of a measuring device on the experiment. The
consideration of these facts led, in thermodynamics, to a new one-parameter family of ideal gases.
In turn, this leads to a new approach to probability theory (including the new notion of independent
events). As applied to economics, this gives the correction, based on Friedman's rule, to Irving
Fisher's ``Main Law of Economics'' and enables us to consider the theory of debt crisis.
\end{abstract}

\section*{Introduction}

The outstanding physicist Ya.~I.~Frenkel wrote:
``We easily get used to the monotonous and unchanging, we stop noticing it.
What we are used to seems natural to us, things we are not used
to seem unnatural and non-understandable.   ...
Essentially, we are unable to understand,
we can only {\it get used to}''\footnote{B.~Ya.~Frenkel,
\textit{Yakov Il'ich Frenkel } (Nauka Publ., Moscow--Leningrad, 1966)
[in Russian], p.~63.}.

In~\cite{1}, Henri Poincar\'{e},  in particular, writes:
``If a physicist finds a contradiction between two theories that are equally dear to him,
he will sometimes say: do not worry about this; the intermediate links of the chain
may be hidden from us, but we will strongly hold onto its ends''~\cite[p.~104]{1}.

Beginning with the creation of
satellites and experiments in the absence of the  gravitational field of Earth, a new
period in physical experimental investigations of thermodynamical phenomena
began.  For example, in an equilibrium state, the liquid
will have the form of a spherical drop.

It should be noted that previously
the relevant experiments were carried out on the surface
of the earth, and hence were subjected to gravitational attraction.
Therefore, the coagulating drops fell to the ground
and the liquid was underneath and the gas above.
Coagulation of drops occurs, in particular, because of the Earth's gravity.

On the other hand, computer-aided experiments have been developed so greatly
that a new science arose, the so-called molecular dynamics.

Significant changes also occurred in the  mathematical sciences.
Therefore, it is not surprising that  great progress
was also made in such a science as thermodynamics.
The difficulty is that everybody is used to the old thermodynamics
based on the Boltzmann distribution.
Mathematical theorems imply some other distributions,
and no contradictions are admissible in mathematics.
Therefore, one should not hold on to the end corresponding
to the old thermodynamics based on the Boltzmann distribution.
As for the Gibbs distribution for the Gibbs ensemble,
this distribution can be justified rigorously\footnote{See~V.~P.~Maslov,
\textit{On refinement  of Several Physical Notions and Solution
of the Problem of Fluids for Supercritical States},
\texttt{\tt arXiv:0912.5011v2 [cond-math.stat-mech], 11~Jan 2010},
Theorem~1.}.

\bigskip

\begin{center}
         * \ * \ *
\end{center}

In his 1903 treatise ``La science et l'hypoth\`ese,'' Henri
Poincar\'e~(\cite{1}, Chap,~11) closely connects probability theory with
problems in thermodynamics. In particular, he writes: ``Has probability been
defined? Can it even be defined? And if it cannot, how can we venture to reason
upon it? The definition, it will be said, is very simple. The probability of an
event is the ratio of the number of cases favorable to the event to the total
number of possible cases\dots We are\dots bound to complete the definition by
saying, `` \dots to the total number of possible cases, provided the cases are
equally probable.'' So we are compelled to define the probable by the probable.
\dots The conclusion which seems to follow from this is that the calculus of
probabilities is a useless science, that the obscure instinct which we call
common-sense, and to which we appeal for the legitimization of our conventions,
must be distrusted.'' \footnote{La probabilit\'e a-t-elle \'et\'e d\'efinie?
Peut-elle m\^eme \^etre d\'efinie? Et, si elle ne peut l'\^etre, comment
ose-t-on en raisonner? La d\'efinition, dira-t-on, est bien simple: la
probabilit\'e d'un \'ev\'enement est le rapport du nombre de cas favorables \`a
cet \'e\'venement au nombre total des cas possibles.\dots On est \dots r\'eduit
\`a compl\'eter cette d\'efinition en disant : ``\dots au nombre total des cas
possibles, pourvu que ces cas soient \'egalement probables.'' Nous voil\`a donc
r\'eduits \`a d\'efinir le probable par le probable.\dots La conclusion qui
semble r\'esulter de tout cela, c'est que le calcul des probabilit\'es est une
science vaine, qu'il faut se d\'efier de cet instinct obscur que nous nommions
bon sens et auquel nous demandions de l\'egitimer nos conventions.}~\cite[pp.~89--90]{1}.

On the other hand, Poincar\'e speaks of the principles of thermodynamics, the
laws of Boyle--Mariotte and Gay--Lussac, and Clausius' approach to molecular
physics. Poincar\'e writes: ``I may also mention the celebrated theory of
errors of observation, to which I shall return later; the kinetic theory of
gases, a well-known hypothesis wherein each gaseous molecule is supposed to
describe an extremely complicated path, but in which, through the effect of
great numbers, the mean phenomena which are all eve observe obey the simple
laws of Mariotte and Gay--Lussac. All these theories are based upon the laws of
great numbers, and the calculus of probabilities would evidently involve them
in its ruin.''~\cite[p.~90]{1}.

Certainly, Poincar\'e gave the standard definition of probability as the ratio
of the number of cases favorable for the event to the total number of possible
events\footnote{``La d\'efinition, dira-t-on, est bien simple: la probabilit\'e
d'un \'ev\'enement est le rapport du nombre de cas favorables \`a cet
\'ev\'enement au nombre total des cas possibles.''}\; and gave a counterexample
to this definition of probability. This definition must be completed, writes
Poincar\'e, by the sentence ``under the assumption that these cases are
equiprobable'' (\cite[p.~90]{1}) and notes that we have completed a vicious
circle place by defining probability via probability.\footnote{``On est donc
r\'eduit \`a compl\'eter cette d\'efinition en disant: `\dots au nombre des cas
possibles, pourvu que ces cas soient \'egalement probables.' Nous voil\`a donc
r\'eduits \`a d\'efinir le probable par le probable.''}

After this, Poincar\'e writes: ``The conclusion which seems to follow from
this\footnote{Poincar\'e presents a series of contradictions in probability
theory, including Bertrand's paradox.}\; is that the calculus of probabilities
is a useless science, that the obscure instinct which we call common sense, and
to which we appeal for the legitimization of our conventions, must be
distrusted.'' (\cite[p.~89]{1}).\footnote{``La conclusion qui semble r\'esulter
de tout cela, c'est que le calcul des probabilit\'es est une science vaine,
qu'il faut se d\'efier de cet instinct obscur que nous nommions bon sens et
auquel nous demandions de l\'egitimer nos conventions.''}

Thus, the problem is to define first of all what cases are to be regarded as
equiprobable in the most natural way. ``We are to look for a mathematical
thought,'' writes Poincar\'e, ``where it remains pure, i.e., in arithmetic''
(\cite[p.~8]{1}). \footnote{``Il nous faut chercher la pens\'ee math\'ematique
l\`a o\`u elle est rest\'ee pure, c'est-\`a-dire en arithm\'etique.''}

Kolmogorov's definition of elementary events which can be taken for a
``complete family of equiprobable events'' is based on intuition, i.e., on
``the obscure instinct which we call common-sense,'' see above. Kolmogorov's
theory, and especially the concept of independent events, well agrees with the
Boltzmann distribution. Kolmogorov's probability theory gave way to a wide
spectrum of applications.

A new conception of thermodynamics, which differs from the Boltzmann
distribution, must lead to a new conception of probability theory, and
especially to a new interpretation of the notion of independent variables.

Thus, Poincar\'e, a great mathematician, who became the Chairman of the Theory
of Probability at the University of Paris (the Sorbonne) at the young age of
32, considers the possibility ``that the calculus of probabilities is a useless
science'' and rejects the idea that ``this calculus'' is to ``be condemned,''
in particular, because of thermodynamics (and ``the kinetic theory of gases'').
Since we have constructed a new thermodynamics, the idea arises to construct a
new probability theory corresponding to this thermodynamics in such a way that
the notion of independent events would correspond to the new Kolmogorov's
conception, namely, to complexity theory, rather than the old notion
corresponding to the Boltzmann distribution.

First of all, in Kolmogorov's theory, the passage to the limit with respect to
the number $N$ of independent tests is of the form
$$
p_i=\frac{N_i}N,
$$
where $N_i$ stands for the number of cases favorable for the event, as
$N\to\infty$ and $\sum N_i=N$. And, if the expectation $M$ of the family
$\{\lambda_i\}$ is given before the passage to the limit with respect to the
number of tests, then we can write
\begin{equation}
MN=\sum\lambda_iN_i, \qquad \sum N_i=N. 
\end{equation}
Denote $MN$ by $M'$. Let us take into account that the numbers $M'$ and $N$ are
large.

Consider the simplest case $\lambda_i=i$,
\begin{equation}
M'=\sum iN_i, \qquad \sum N_i=N. 
\end{equation}
I hope there will be no confusion if we shall omit the prime.
\medskip

\begin{example}
In the problem known under the title ``partitio
numerorum''~\thetag{2} (see Example~5 below), we may consider all versions of
partitioning a number~$M$ into~$N$ summands as equiprobable events, provided
that $N<M$. For example, we may assume that the partitions of the number 5 into
the sum 1+4 and into the sum 2+3 are equiprobable.
\end{example}

Here we undertake an essential deviation from the Kolmogorov probability;
however, we approach Poincar\'e's point of view concerning arithmetic. The
question is, what are specific problems to which the new probability theory can
be applied?

Kolmogorov~\cite{2} writes: ``The probability approach is natural in the theory
of transmission, over communication channels, of ``mass'' information
consisting of many disconnected or weakly connected messages subjected to
certain probability laws. In problems of this kind, the confusion of
probabilities and frequencies within the limits of a single sufficiently long
time series (this confusion can rigorously be justified under the conjecture of
a rather fact mixing), which is deep-rooted in applied researches, is also
practically harmless.'' \dots``If there is still some dissatisfaction, it is
related to a certain vagueness of our conceptions dealing with relationships
between the mathematical probability theory and practical ``random phenomena''
in general.''

The new probability theory corresponding to the new conception of
thermodynamics must adequately describe situations (for example, in computer
simulation) with many agents, and also for the case in which the analysis
involves ``people.'' This theory must meet the requirements of semiotics when
the number of symbols is very large. This theory must be related to the Zipf
law and to similar laws, and also must be applicable to stock exchange
speculation (see~\cite{3}). As we shall see below, the velocity of money and
the occurrence and repayment of debts are also related to problems of new
probability theory.

Introduction of an additional parameter~$N$ in the probability theory
is very essential, although it finally tends to infinity, $N\to\infty$.
In particular, it follows from the principle of Lagrange undetermined multipliers
that, in the process of calculation of the entropy maximum,\footnote{For the definition
of Bose entropy, see~\cite{11}, the section
``Nonequilibrium Fermi and Bose gases'' in the three-dimensional case;
the general case is considered in Section~1.1
of the present paper.}
the quantities $N$ and $M$ are associated with the following two
parameters: the chemical potential~$\mu$
and the inverse temperature $\beta=1/T$, where~$T$ is the temperature.
Thus, the ``number of measurements''~$N$ is associated with a new parameter~$\mu$,
which must play a key role in the new probability theory.

\bigskip

\section{G\"odel's theory, ``sand pile'' paradox, and the Bose--Einstein
distribution}

{{\hskip3.6truecm Albert Einstein remarked towards the end of his career that he
only\hfill

\hskip3truecm went to his office at Princeton ``just to have the privilege of
walking\hfill

\hskip3truecm home with Kurt G\"{o}del.''\hfill

\hskip4truecm``{\it The Forgotten Legacy of G\"{o}del and Einstein\;}''
P.~Yourgrau}

}

\bigskip

There is a famous dictum of Kronecker, one of the greatest mathematicians of
the 19th century: {\it``The natural numbers are from God, all the rest is the
handiwork of man.''} In the paper ``On the Dogma of the Natural Numbers''
[Uspekhi Mat. Nauk, {\bf28} (4), 243--246 (1973), in Russian], P.~K.~Rashevskii
[Rashevsky] wrote: : ``The famous negative results of G\"odel in the thirties
are founded on the belief that, however long you continue the construction of
mathematical formulas for a given (totally formalized) mathematical theory, the
principles of counting and ordering the formulas remain ordinary, i.e.,
subjected to the scheme of natural series. Certainly, this belief was not even
explicitly stipulated, because it was assumed to be obvious to this very
extent.''

Recall that G\"odel's ``negative results'' (the impossibility of proving the
consistency of arithmetic by means of arithmetic or any formal theory
containing arithmetic) destroyed the foundation of Hilbert's program for
constructing all of mathematics as a completely formalized system. Indeed, it
follows from G\"odel's Incompleteness Theorem that any formal system containing
the natural number series (i.e., containing its usual axiomatization) is flawed
in principle: if it is consistent, it must be incomplete, since it must contain
true arithmetical statements that can neither be formally proved nor disproved.
The analysis of the constructions of the great G\"odel shows that these
arithmetical statements are related, via the principle of mathematical
induction, to extremely large numbers. It is well known that Hilbert, after the
incompleteness of any noncontradictory formal theory containing arithmetic was
proved, fell into a prolonged depression.

The author conjectures that G\"odel's theorem leads to the explanation of
several paradoxes of thermodynamics, provided one makes use of an additional
instrument, for instance, of a measuring device. Let us present a few
elementary examples.

Already in the 4th century B. C., the Greek philosopher Eubulides of Miletus
formulated the sandpile paradox: beginning with what number of grains of sand
can their collection be regarded as a pile?

First let us comment on this question.

1. When a pile of sand is measured, instead of counting the grains, one
ordinarily uses a spoon, or a cup, or some weighting device,
and any other arithmetic averaging as well.

2. When such a measurement is performed, precision is lost: the spoons are
filled with a number of grains of sand known only approximately, only up to
several grains.

The first factor results in the appearance of a new arithmetic for the subsets
of the given set of grains, i.e., one more mapping of the subsets of the given
set to the set of natural numbers.

The second factor leads to a small loss of precision, to so-called
``fluctuations'' or, from the mathematical point of view, to ``fuzzy sets'' or
``soft sets.'' What fluctuations are admissible?

Consider the simple example of a cup of tea into which we have mixed a spoonful
of sugar. We~wait for the liquid in the cup to stop rotating, to become calm,
but still remain hot. This means that we allow some fluctuation of the liquid;
otherwise, it will cool off. The less fluctuations are allowed, the colder the
tea will become and the longer we will have to wait before beginning to drink
our tea. If the fluctuations are already admissible from our point of view, we
say that the tea has been calmed. (This process is quite similar to passing
from nonequilibrium thermodynamics to equilibrium thermodynamics, in which only
fairly small fluctuations are allowed.)

Already in the second half of the 20th century, mathematicians have attempted
to carry out a {\it reform} of the natural numbers from the point of view of
``fuzzy sets,''\,``interval analysis,''\, ``nonstandard analysis,'' and the
like~\cite{4}--\cite{8}; see also~\cite{9}. To use the language of physics, this
involves taking fluctuations into consideration.

One can modify Kronecker's statement in the following way: {\it God created
air}, and therefore created the Avogadro and the Loschmidt numbers, i.e.,
quantities from $10^{19}$ to $10^{23}$, and also the unavoidable fluctuations
of air. This implies the practical impossibility of calculating such quantities
exactly.

In his famous work ``What is life (from the point of view of physics)?''
Schr\"odinger put forward the hypothesis that if the number of particles $N$ is
large enough, then we cannot obtain a calculation result with precision better
than $\sqrt N$. And Schr\"odinger called this a ``law of nature.'' More
precisely, in mathematical terms, this can be expressed by saying that the
probability of an error in the calculation of~$N$ being greater than $\sqrt N$
is sufficiently small. We call this a ``soft set'' of~$N$ elements.
Schr\"odinger's ``law of nature'' contradicts the dogma (axioms) of the natural
numbers.

Thus, first of all, a sandpile is a soft set, with the important properties
that (1) it presents another mapping (of its subsets) to the natural numbers
and (2) it possesses admissible fluctuations.

Finally, let us pass to the main property of the sandpile, which consists in
the following. If we interchange two grains of sand in the pile, then it
remains the same pile, i.e., the concept will not change and the rules of its
arithmetic will not change.

This means that although the grains of sand are different (they can be
distinguished by a scrupulous study), within the sandpile they lose their
individuality. And therefore, in the pile they do not obey the Boltzmann
statistics, but a statistics of the Bose--Einstein type.

For example, if we buy a pound of sugar, the calculation of the exact number of
grains of sugar is not only extremely difficult, but also has no practical
meaning. However our ``pile'' of sugar can be divided into two half kilo piles
or into ten piles of 100 grams each. The addition and multiplication for them
remain. In other words, for piles the rules of arithmetic still hold. The scale
is determined by a macro-measuring device (for example, by a balance with
weights; we assume for the sake of simplicity of our presentation that there
is also a minimal weight).

Let us repeat once again: if, under the interchange of two grains, the notion
of pile does not change, and neither do the rules of arithmetic for piles, then
the individuality of particles of sand within the pile are lost. This is the
main property of piles. From the point of view of statistical physics, this
corresponds to passing from Boltzmann statistics (from the Boltzmann entropy,
which is equal to the Shannon entropy) to a one-parameter family of statistics
of the Bose--Einstein type, although grains of sand {\it can be distinguished
from each other} and there are no identical grains.
\medskip

\begin{example}\label{exam2}
Let us put two one-copeck coins in two banks
and calculate the number of possible variants
of this decomposition.

(1) We put two coins in the first bank and nothing in the second one.

(2) We put two coins in the second bank and nothing in the first one.

(3) We put per one coin in each bank.

Thus, we have three variants. This is the Bose-type  statistics\footnote{L.~Landau and E.~Lifshits  explain
the identity principle for particles as follows: : ``In
classical mechanics, identical particles (such as electrons) do not lose their
'identity' despite the identity of their physical properties. ... we can
'number' them and then observe the motion of each of them along its
trajectory; hence, at any instant of time, the particles can be identified ...
In quantum mechanics, it is not possible, in principle, to observe each of the
identical particles and thus distinguish them. We can say that, in quantum
mechanics, identical particles completely lose their 'identity'\,''
\cite{82}, p.~252. We say that these particles are objectively indistinguishable. The elements of  a pile can be enumerated and we say that
its elements are subjectively indistinguishable  (as coins in Example~\ref{exam2}).}.

Now we calculate the number of possible variants
of putting a one-copeck coin and a one-pence coin in two banks.
In this case, variant~(3) splits into the following two variants:
we put the copeck in the first bank and the pence in the second bank
or we put the pence in the first bank and the copeck in the second bank.
As a result, we have four possible variants.
This is the Boltzmann statistics.

Similarly, the decomposition of $M$ copeck coins over $N$ banks
gives the Bose-type statistics.
If you put a 1000-rouble bond on your bank account
and then want to take it back, it is hardly probable
that you obtain the same 1000-rouble bond (with the same number).

Quite similarly, the conveyance of $M$ loaves of bread from a big baker's shop
over $N$ bakeries satisfies the Bose-type statistics.
But the variety of proposed loaves,
the purchaser in a bakery can choose a certain loaf that he likes most of all.
These  situations are determined by the words ''wholesale''
(i.e. vegetables and
fruit are measured  in containers, e.g. in barrels with  volume  $V_\delta$)
and ``at retail''.

This is an example of the ``sandpile'' paradox:
while the set of elements is treated as a pile,
its elements are subjectively indistinguishable and satisfy the Bose-type  statistics.
In our case, it is meaningful to distinguish the elements of the set,
then this is already not a pile, and it satisfies the Boltzmann statistics.
\end{example}

If we consider a set of grains before the formation of a pile (i.e., before the
time moment at which the set can be measured by a macroscopic device), then the
transposition of grains gives another natural series as compared with the
original one. If we continue extending the new natural number series according
to Boltzmann, the difference between the statistics will disappear. This is
similar to the one-parameter family of hyperbolic geometries whose curvature
vanishes as they become Euclidean.

Let us cover the actual (practical) natural series\;\footnote{This means that
all natural numbers are imagined as existing simultaneously.} by a system of
soft subsets of $N$ elements, which meet one another in the domain of
fluctuations in general. It is assumed that the union of these subsets contains
the entire natural series and that $N>10^6$. The union of pairs of subsets of
$2N$ elements contains the natural series again and, continuing the process of
adding a new pile, $3N$, $4N$, \dots, we obtain an arithmetic of soft subsets
and a practical natural series of piles.

We would like to say in advance that equilibrium thermodynamics is related to
these very deep logical problems.

If we supplement the dogma of the natural numbers with an external measuring
device without denying the commutativity of addition, this will not contradict
the fact that we are counting only up to some precision, and if probability is
taken into consideration, will allow counting up to some given ``soft''
precision. This soft precision allows to count all the elements of a finite
set. We are only saying that if this unprecise count differs from the exact
one, then we can disregard the error. The so-called equal distribution law only
concerns small natural numbers. This law must also be modified. And this
contradicts our habitual philosophy, accepted for centuries, and can therefore
generate protests not only from scientists, but also from philosophers.

Progressive Russian men of letters, for example, Chernyshevskii, \footnote{We
will not quote the adjectives with which Chernyshevskii crowned the great
mathematician Lobachevsky in his ``Letters to my sons A. N. Chernyshevskii and
M. N. Chernyshevskii'' in 1878.} categorically refused to accept Lobachevsky's
geometry.

Note that Lobachevskian geometry is in fact a one-parameter family of
geometries, depending on the radius (curvature), which passes to Euclidean
geometry as the radius tends to infinity.

Soviet philosophers ostracized the followers of Bohr's Complementarity
Principle, which he borrowed from biology and psychology. At the same time, if
we accept this complementarity principle as a ``complement'' to the new concept
of the number $N$ of particles with admissible fluctuations of $\sqrt N$, then
we come to the notion of chemical potential~$\mu$ and to an assertion similar
to Heisenberg's Indeterminacy Principle: the smaller the fluctuations of~$\mu$,
the smaller the fluctuations of~$N$. And, further, we come to other intensive
and extensive quantities that characterize thermodynamics.

First of all, the pressure $P$ exercised by the piston decreases the volume $V$
of air in the vessel, and Bohr's Complementarity Principle relates these two
substances. The increase of the temperature $T$ increases the chaotic speed of
particles, it increases the chaos determined by the entropy. These two
quantities correspond by Bohr's Complementarity Principle.

Bohr's Complementarity Principle is related to Heisenberg's Indeterminacy
Principle, namely, the decrease in the fluctuation of one component corresponds
to an increase of the complementary quantity. {\it Bohr explains this principle
by the interference of the measuring device.} In our example with sugar, the
macro-measuring device (the balance) also plays its role. Especially if, in the
weighting process, ``self-feeding'' processes occur, similar to to those taking
place when temperature is measured by a mercury thermometer, which ``absorbs
into itself'' part of the energy (heat) of the particles that it is supposed to
measure.

Bohr's Complementarity Principle was developed and applied not only by
physicists with a wide outlook such as Max Born, but also by such pragmatists
as Pauli. According to the Complementarity Principle, motion and immobility,
for example, were compared. But a special role is played by the Complementarity
Principle in equilibrium or nonequilibrium thermodynamics.

Just as the existing ``time-energy'' complementarity, in thermodynamics there
is the complementarity ``observation time -- size of the fluctuation,'' since
we come to a situation of equilibrium thermodynamics, as we explained in the
cup of tea example. Here an essential role is played by viscosity, the slowing
down phenomenon that leads to equilibrium thermodynamics {\it up to a concrete
value of the fluctuation}. The old thermodynamics was based on the Boltzmann
statistics, and the latter on the dogma of natural numbers. But it turned out
that the dogma of natural numbers is just as unfit for very large collections
considered in thermodynamics as Euclidean geometry is for the description of
the Universe.

Below we construct a one-parameter family of ideal gases that describe such
very large collections.

\begin{example}
As an example of the famous Erd\H os theorem from number
theory, we consider the solution of the ancient problem {\it partitio
numerorum}. This problem features an integer~$M$, which is expanded into $N$
summands; for example, suppose that $M=5$ and $N=2$:
$$
5=1+4=2+3;
$$
this yields $\mathcal{M}=2$ variants of the solution of this problem.

If $M=10^{23}$ and $N=1$, then there is only one variant of the expansion:
$\mathcal M=1$. If $M=10^{23}$ and $N=10^{23}$, then there is also only one
variant of the expansion: the sum of 1's, i.e., $\mathcal M=1$.

Obviously, for a fixed number~$M$, there exists a number~$N_c$ for which the
number of variants~$\mathcal M$ of the expansion is a maximum (in general, this
number is not unique). The quantity $\log_2\mathcal M$ is called
{\it Hartley's entropy}. At the point where it attains its maximum, one has
the maximum of the entropy. The chemical potential equal to zero
corresponds to this point.

Suppose we are given the partition
$$
M=a_1+\dotsb+a_N
$$
of the number~$M$ into $N$ summands. Let $N_j$ be the number of summands on the
right-hand side of this equation that are precisely equal to the number~$j$.

Then there are $\sum_jN_j$ in all, and this number is equal to~$N$, because we
know that there are only $N$ summands. Further, the sum of summands equal
to~$j$ is $jN_j$, because their number is $N_j$; hence the sum of all summands
is obtained by summing these expressions over~$j$, i.e., $\sum_jjN_j$, and it
is~$M$. Namely,
\begin{equation}
\sum_0^\infty N_i=N, \qquad \sum_0^\infty iN_i=M. 
\end{equation}

The nonuniqueness of this maximum and the indeterminacy of the number of these
maxima allowed Erd\H os to obtain his result only up to $o(\sqrt M)$. Thus, he
retreated from the dogma of the natural numbers in the direction of the notion
of soft set. \footnote{The principal term $N_{\mathrm{cr}}$ for which the
number of solutions $\mathcal M$ of system~\thetag{3} is maximum is of the form
$c\sqrt M\ln M$, where $c$ is a strictly defined constant.}
\end{example}

\medskip

Relations~\thetag{3} correspond to physical relations of the form
\begin{equation}
\sum^\infty_{i=0} N_i=N, \qquad \sum^\infty_{i=0}\varepsilon_i N_i={\mathcal E}, 
\end{equation}
where $N_i$ is the number of particles at the $i$th energy level, the
$\varepsilon_i$ are the discrete collections of energies, and $\mathcal E$ is
the energy.

The derivation for the general case, which coincides for $E=p^2/2m$ (where $p$
stands for the momentum and $m$ for the mass of the particle), coincides with
the Bose distribution presented for the volume $V=1$ by Landau and Lifshits
in~\cite{11}, is considered in~\cite{12--14}.

\subsection{One-Parameter Family of Distributions of the ``Partitio
Numerorum'' Type}

Let us briefly present results of the papers listed above.

If we consider the particular nonrelativistic case in which the Hamiltonian~$H$
of the system is ${p^2}/{2m}$, where $p$ is the momentum and~$m$ the mass,
then, up to constant multipliers, problem~\thetag{3} corresponds to the
two-dimensional case.

Consider a straight line and a plane. Let us mark off the points
$i=0,1,2,\dots$ on the straight line and the points $x=i=0,1,2,\dots$ and
$y=j=0,1,2,\dots$ on the coordinate axes $x,y$ of the plane. With this set of
points $(i,j)$ we associate the points on the line, which constitute the
sequence of natural numbers $l=1,2\dots$. To each point let us assign the pair
of points $i$ and $j$ by the rule $i+j=l$. The number of such points $n_l$ is
$l+1$. This is the two-dimensional case.

Consider the three-dimensional case. On the axis, set $z=k=0,1,2,\dots$, i.e.,
set $i+j+k=l$. In this case, the number of points $n_l$ is
$$
n_l=\frac{(l+1)(l+2)}2.
$$

For the $D$-dimensional case, it is easy to verify that the sequence of weights
(multiplicities) of the number of variants
$$
\text{ $i=\sum_{k=1}^Dm_k$,}
$$
where the $m_k$ are arbitrary natural numbers, is of the form:
\begin{equation}
q_i(D)=\frac{(i+D-2)!}{(i-1)!(D-1)!}. 
\end{equation}

The three-dimensional case $D=3$ corresponds to the following problem of number
theory (see~\cite{11}):
\begin{equation}
\sum N_i=N, \qquad \varepsilon\sum\frac{(i+2)!}{i!6}N_i={\mathcal E}, \qquad
\frac{\mathcal E}{\varepsilon}=M. 
\end{equation}

As was already stated, Schr\"odinger thought that the statistical laws valid as
$N\to\infty$, where $N$ is the number of particles, hold with accuracy at most
up to~$\sqrt N$.

However, such a rough estimate also has a positive aspect.

This consideration enables us to extend the number theory presented above no
noninteger dimensions.\footnote{The fractional dimension in number theory has
nothing in common with the dimension in the three-dimensional configuration
space. This fractional dimension is related only to the spectral density of a
separate molecule and can be expressed in terms of the fractional dimension of
the momenta, $\{p\}$, if it is assumed that the energy is proportional to
$p^2$. The Schr\"odinger equation for a separate molecule has a rather
complicated spectrum. The fractional dimension in the momentum space
corresponds to some averaged density of this spectrum and is a macroscopic
quantity which can be measured experimentally by using the dimensionless
quantity $Z$ at the critical point.}

Let us consider expressions of the form
\begin{equation}
\varepsilon\sum\frac{\Gamma(d+i)}{\Gamma(i+1)\Gamma(d+1)}N_i\le {\mathcal E}, 
\end{equation}
\begin{equation}
\sum N_i=N, 
\end{equation}
where $d=D/2$ and $\Gamma(d)$ is the gamma function.

Let $\mathcal M$ be the number of solutions satisfying inequality~\thetag{7}
and relation~\thetag{8} for a noninteger~$D$ (the ``fractal dimension'').

Let us define the constants:  the chemical potential $\mu$ and  the inverse temperature  $\beta=1/T$
``conjugate''\;\footnote{Conjugacy is understood in the sense of the Lagrange method of
multipliers (cf.~Bohr's Complementarity Principle discussed above).} to~$N$
and~${\mathcal E}$.

Consider the three-dimensional case. We have
\begin{equation}
N=\sum^\infty_{i=1}\frac{(i+1)(i+2)}{2(e^{\beta(i-\mu)}-1)}=\sum N_i, 
\end{equation}
\begin{equation}
M=\frac{\mathcal E}{\varepsilon}
=\sum^\infty_{i=1}\frac{i(i+1)(i+2)}{2(e^{\beta(i-\mu)}-1)}. 
\end{equation}

This is a well-known result in number theory~\cite{12}--\cite{14}.
It is very simple  as compared with the Erd\"os theorem.

Here we have used the fact that arithmetic summands can be rearranged and have
summed the elements of the ``pile'' over all rearrangements each corresponding
to a particular sequence of natural numbers. According to Boltzmann, each
rearrangement yields a new pile, but, for us, it is one pile. Thus, we carry
out a procedure similar to that used by Landau and Lifshits in their
calculation of the Bose--Einstein distribution.

Now the distribution takes the form of the Bose--Einstein
distribution~\cite{11} in which $V=1$.

In the general case, for $D>2$,
\begin{equation}
{\mathcal E}=\varepsilon\sum^\infty_{i=1}\frac{iq_i}{e^{\beta(i-\mu)}-1}, 
\end{equation}
where
\begin{equation}
q_i=\frac{\Gamma(i+d)}{\Gamma(d)\Gamma(i+1)},\qquad
N=\sum^\infty_{i=1}\frac{q_i}{e^{\beta(i-\mu)}-1}. 
\end{equation}

Since, for a fixed~$k$, the number of particles $N$ tends to infinity, we can
pass from sums to integrals by the Euler--Maclaurin estimates and obtain the
following relation.

Landau and Lifshits~\cite{11} write: ``Any macroscopic state of an ideal gas
can be characterized as follows. Let us distribute all the quantum states of an
isolated particle of the gas over groups each of which contains close states
(possessing, in particular, close energies); further, the number of states in
each group and the number of particles contained in them is very large.
Renumber these groups of states by $j=1,2,\dots\,$, and suppose that $G_j$ is
the number of states in the $j$th group, while $N_j$ is the number of particles
in these states. Then the collection of numbers $N_j$ will fully characterize
the macroscopic state of the gas''~\cite[Russian, p.~143]{11}.

The ``groups,'' which are also called ``cells,'' contain the quantum
eigenvalues (states). The number of these states in each cell can be roughly
associated with the quantum spectrum density and corresponds to the fractal
dimension.

Therefore, without regard for the interaction, we obtain a macro parameter
defining the spectrum density $\widetilde Z=\mathcal E/NT$ of a given molecule.
We can state that this quantity $\widetilde Z$ is the macro measurement of the
whole quantum spectrum density of a given molecule  for the case
of  the maximal $\mathcal{M}$, i.e.  for the chemical potential  $\mu =0 $.

Landau and~Lifshits make the following remark concerning the ``momentum
space:'' ``The phenomenon of accumulation of particles at the state with
$\varepsilon_i$, $i=0$  is often referred to as the ``Bose--Einstein condensation.'' We
stress that one can speak here only about the ``condensation in the momentum
space;'' {\it certainly $(!)$, there is no practical condensation in the
gas}''~\cite{11}. However, we claim that a practical condensation in an
ordinary gas occurs indeed under certain conditions, see Section~2.

The energy $\mathcal E$ (which, when divided by $\varepsilon$, is equal to
the original number $M$ in the problem of {\it partitio numerorum\/})
for $\gamma=(d-1)/2$, where $d$ stands for the ``fractal'' noninteger
dimension, $d=D/2$, is of the form
\begin{equation}
\mathcal E=\frac{T^{2+\gamma}}{\Lambda^{2(1+\gamma)}\Gamma(2+\gamma)}
\int_0^\infty\xi^{1+\gamma}\bigg\{\frac1{e^{(\xi-\kappa)}-1} \bigg\}\,d\xi
=\frac1{\Lambda^{2(1+\gamma)}}T^{2+\gamma}\operatorname{Li}_{\gamma+2}(e^{\mu/T}),
\qquad \kappa=\frac\mu T, 
\end{equation}
where $\operatorname{Li}_\gamma(\cdot)$ is the polylogarithm,
$\operatorname{Li}_{\gamma+2}(1)=\zeta(\gamma+2)$, e.g.,
for $\mu=0$, $\zeta$ is the Riemann zeta function, and $\Lambda$
is a constant.\ \footnote{The relation of Hartley's entropy equal to
$\ln\mathcal M$ with the quantity $\partial\mathcal E/\partial T$ is established by a direct
calculation~\cite{17}.}

Now we can rigorously mathematically formulate the main principle of
thermodynamics corresponding to the approximate conservation of the gas density
(this corresponds to the physicists' statement in equilibrium thermodynamics:
``the density is homogeneous in a vessel;'' physicists consider equilibrium
thermodynamics as a separate discipline, after which fluctuation theory is also
considered separately).

The experimenter assumes that the density of particles inside the vessel is
constant up to fluctuations (of the order of the root of the number of
particles inside a small subvolume). This means that he counts the number of
particles in each small subvolume in which there are at least one million
particles and the density is approximately constant. In each fixed subvolume
$V$, there is the same fuzzy number of particles~$N_\delta$. Obviously, any
rearrangement of the numbers of particles does not affect their density (just
as in arithmetic, a rearrangement of summands does not affect their sum). If we
even assume that the experimenter numbered all the particles in the previous
measurement, then, in the next measurement, e.g., after increasing the pressure
inside the vessel by using a piston and after achieving the equilibrium (i.e.,
the homogeneity of the density in the vessel), he cannot state what number
corresponds to a particular particle and, therefore, must again number all the
particles to find their density. This is a fact of arithmetic, and arithmetic
is the foundation of analytic number theory.

\bigskip

\subsection{The Main Principle of Equilibrium Thermodynamics {\rm(}Uniform
Density{\rm)} and the Family of Ideal Gases Corresponding to Every Molecule}

First, let us roughly state this principle: the density (more precisely,
concentration) of a gas in a closed vessel is almost constant and almost equal
to the number of particles in the vessel divided by the volume of the vessel.

This principle can be stated in a mathematically rigorous way as follows.

We consider molecules of one type, i.e., of the same spectrum density.
As was already shown, the spectrum density is characterized by the
parameter~$\gamma$.
Assume that the spectrum density of the molecules of this type
corresponds to the parameter $\gamma=\gamma_0$.

\medskip

{\it Consider a vessel of volume $V$ containing $N>10^{19}$ identical molecules
corresponding to the parameter $\gamma=\gamma_0$. Consider a small convex
volume of size~$V_\delta$ containing $N_\delta$ particles, where $N_\delta$ is
not less than $10^6$. Let $\mathbb P$ be the probability of the event
consisting in the deviation of $N_\delta$ by a value exceeding
$\sqrt{N_\delta}$ for any volume of size~$V_\delta$ inside the vessel. The main
principle is that the probability~$\mathbb P$ is sufficiently small.}

\begin{remark} This obvious fact strictly implies a new relation for the
ideal gas, separate for each molecule of a pure gas. This is a consequence of a
sufficiently simple theorem of number theory.
\end{remark}

\begin{remark} The ratio $M/{kNT}$ is ``dimensionless'' (owing to the
Boltzmann constant~$k$) in the~sense that, in physics, $M$ corresponds to the
energy $\mathcal E$. The quantity $Z={\mathcal E}/{\mathcal RT}$, where
$\mathcal R$ is the gas constant, is called the {\it compressibility factor}.
\end{remark}

It follows from this principle that the numbering of the particles in a
subvolume $V_\delta$ is arbitrary, and the concentration (density) is
independent of it. We can rearrange the numbers, which does not lead to any
change in density: a rearrangement of summands does not affect their sum.
Combining this with \thetag{13}, we prove the following statement.

\begin{corollary} At $\mu=0$, the maximum of $\mathcal M$ and the entropy
is attained. We obtain a one-parameter family {\rm(}with parameter~$\gamma${\rm)}
of maximum values of critical compressibility factors
\begin{equation}
Z_{\mathrm{degen}}=\frac
M{kNT}\bigg|_{\mu=0}=\frac{\zeta(\gamma+2)}{\zeta(\gamma+1)}. 
\end{equation}
\end{corollary}

What happens with a pure gas for $T<T_{\mathrm{degen}}$? We refer
relation~\thetag{13} for any $\gamma$ as the ``Bose gas,'' always using the
inverted commas in this case.

We see that the formulas obtained above coincide for $\gamma=3/2$ with the
formulas for the Bose gas. Landau and Lifshits especially warn in the quatation
given above that one must not confuse the Bose condensate with a practical
condensate. However, we should ask: Where are the excessive particles when the
temperature became lower the ``degeneration temperature'' for the pure gas for
which we have rigorously developed the above formulas~\thetag{13} from the main
postulate? Despite the rigid taboo claimed by Landau and Lifshits (see above),
it is natural to assume that the excessive particles had condensed into the
liquid phase.\footnote{And, preliminarily, into dimers. Van der Waals said in
1900 that his model is inaccurate because it ``does not consider associations
of molecules,'' i.e., the formation of dimers.}

We shall prove this rigorously in the second section for the Wiener
quantization of thermodynamics. For now, we present a comparison of
experimental graphs with graphs constructed according to our computations and
for diverse pure gases in the paper by Apfel'baum and Vorob'ev ``Correspondence
between of the ideal Bose gas in a space of fractional dimension and a dense
nonideal gas according to Maslov scheme''~\cite{18}.

Thus, we assume that the critical value of the compressibility factor
$Z_{\mathrm c}$ for the given gas is of the form
\begin{equation}
Z_{\mathrm c}=\frac{\mathcal E_{\mathrm c}}{{\mathcal R}T_{\mathrm c}}
=\frac{\zeta(\gamma+2)}{\zeta(\gamma+1)}, \qquad \mathcal E_{\mathrm
c}=P_{\mathrm c}V_{\mathrm c}. 
\end{equation}
Hence we obtain the parameter $\gamma$ characterizing the spectrum (``spectral
density'') of a given molecule. \footnote{In general, we cannot divide the
internal energy $\mathcal E_{\mathrm c}$ (obtained from the spectrum density)
by the product $P_{\mathrm c}\times V_{\mathrm c}$. But, at the critical point,
their experimental values are known and we shall use them.}

Let us compare the critical isotherms on the $V,P$ plane up to the critical
point (at which the density is sufficiently high and the pressure is 3--4 times
greater than that of the ``ordinary'' Boltzmann ideal gas).

\begin{figure}[h]
\begin{center}
\includegraphics{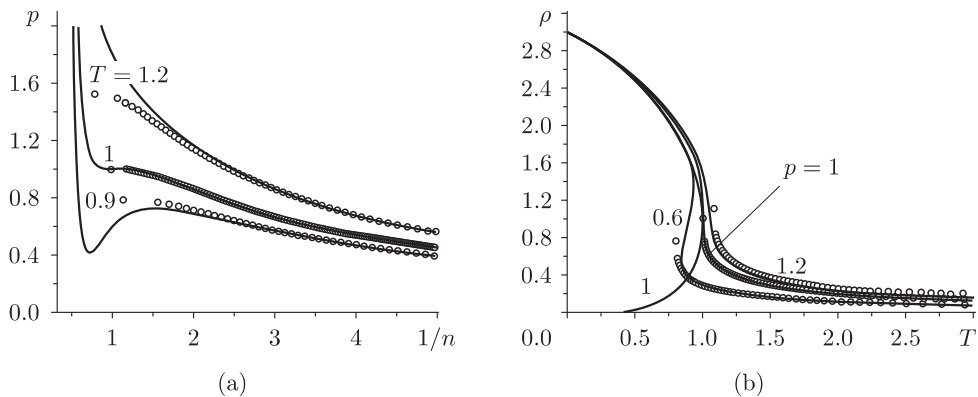}
\end{center}
\caption{(a) Isotherms of pressure for the van der Waals
equation are shown by continuous lines. The small circles show the
corresponding lines computed with $\gamma=0.312$ for $\varphi(V)=V$
(i.e., the ideal ``Bose gas''), $Z_{\text{cr}}=3/8$, $p=P/P_{\text c}$, and $n=N/N_{\text c}$.
\quad
(b) Isobars of density for the van der Waals equation are shown by continuous
lines. Line 1 is the binodal curve. The small circles correspond to isobars for
the ``Bose gas'' with $\gamma=0.312$.}
\end{figure}

\begin{figure}[h]
\begin{center}
\includegraphics{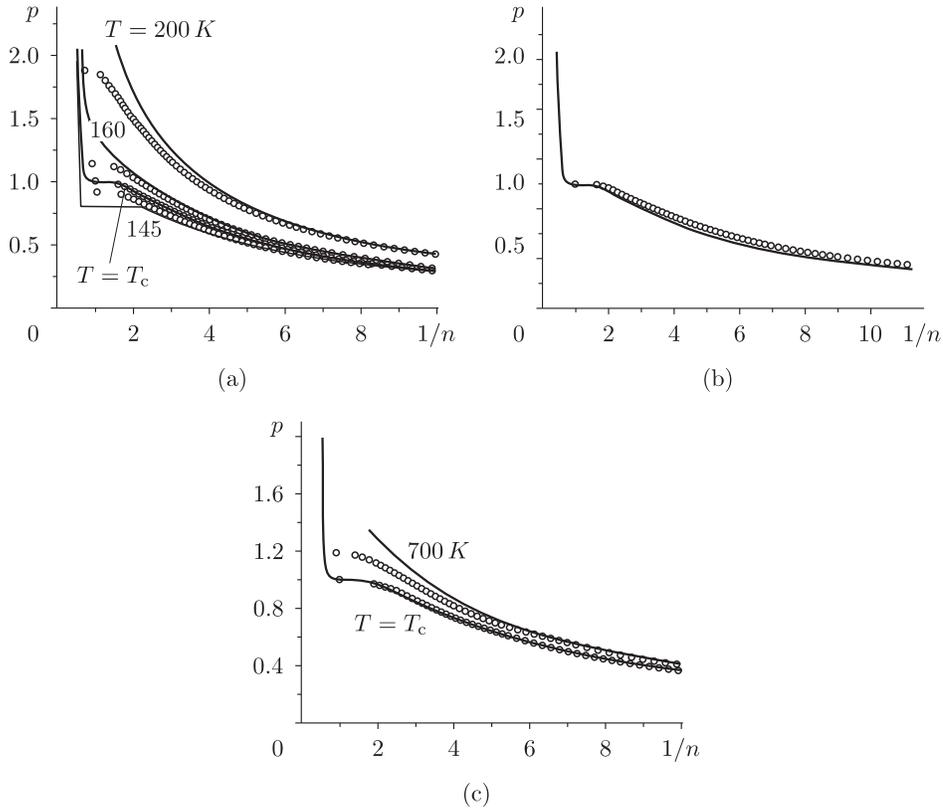}
\end{center}
\caption{a) Isotherms for argon. The continuous lines
correspond to the experimental data, and the line formed by small circles is
constructed according to the isotherm of ideal ``Bose gas,''
$Z_{\text{cr}}=\frac{\zeta(\gamma+2)}{\zeta(\gamma+1)}=0.29$,
$p=P/P_{\text c}$, and $n=N/N_{\text c}$.\quad
(b) The same for water, $Z_{\text{cr}}=0.23$.\quad
(c) The same for copper, $Z_{\text{cr}}=0.39$.}
\end{figure}

The approach which takes into account the possibility to preserve the pile
under the transposition of two grains fundamentally solves the Gibbs paradox.
Note that, although the greatest mathematicians, including von Neumann and
Poincar\'e, tried to patch up the Gibbs paradox, they did not succeed. In the
same way, many great physicists tried to clarify the problem with the paradox,
and still, all the time, rather serious works occur that try to coordinate with
the Gibbs paradox. Last year, ten papers concerning the topic were presented to
arXiv.org.

V.~V.~Kozlov~\cite{19} rigorously proved the existence of an entropy jump in
the Poincar\'e model for gas in a parallelepiped with mirror walls, i.e., he
establishes the presence of the paradox in this model. The rigorous
mathematical work~\cite{19} helped to find the solution suggested by the author
of the present paper for the Gibbs paradox.

\subsection{On the table of molecules versus the energy-spectrum density}

The question is how to calculate the average density of the spectrum
of a molecule at a given temperature.
We consider the energy $\mathcal{E}$ corresponding to $N$ particles
in the maximum chaos situation where the number of possible variants
of the system solution is maximal, i.e., at the entropy maximum
(in the case of maximum indeterminacy).
The quantity $\mathcal{E}$ is divisible by $N$;
this is the average energy of a single particle.
Dividing this average energy by the energy $kT$,
we obtain the quantity of energy of a particle per unit energy
corresponding to the given temperature: $Z=\frac{\mathcal{E}}{NkT}$.
In what follows, we omit the Boltzmann constant,
which cannot lead to misunderstanding,
because this only means that the temperature is measured in energy units.

We note that a similar construction is developed
on the basis of the Boltzmann distribution in Sec.~40 in~\cite{11},
which is called ``Nonequilibrium perfect gas''.

In our case, the quantity $\mathcal{E}/NT$ depends on the dimension $\gamma$
in the space of momenta (or energies $E=\frac{p^2}{2m}$, $dE=\frac{p\,dp}{m}$)
and hence is a one-parameter family (depending on the parameter~$\gamma$,
which is the Hausdorf dimension, fractional or fractal dimension).

We obtain a table arranged not according to masses
as the Mendeleev periodic table but according to the average energy density.
This table comprises all known molecules.
The molecule internal energy can be represented as $PV$.
The extreme point at the maximal entropy corresponds
to the catastrophe in the sense of Arnold.
This is (because of the pile basic property)
the degeneracy point corresponding to the degeneracy point of the Bose gas
in the fractal (fractional) dimension.

Here we do not force the parameter $\gamma$ to fit the experiment.
To say that the parameter $\gamma$ is adjusted to experiments
is the same as to say that the Mendeleev periodic table is forced
to fit the experiment with respect to the mass parameter.

In addition, I note that the maximum entropy point
(the point $\mu=0$) corresponds to the focal point
(or to the point of catastrophe, as it was called by Arnold)
similar to the focus of a lens focusing the Sun rays.

We answer the above question as follows.
The mathematicians have been using the dogma of the natural numbers
for a very long time.
G\"odel disproved it only in 1930, and we have already described the state
to which the greatest mathematician Hilbert was brought by this proof.
Boltzmann died much earlier. He committed suicide,
and it was said that precisely the claims of mathematicians
underly his deed.
As is known, in a very hot discussion with mathematicians about molecules,
he shouted: ``Go and verify them!''.
If he were alive in the after-G\"odel times, he would cry to the mathematicians:
``Go and number them!''.
And the mathematicians who were acquainted with G\"odel's work
could not say anything against this.

\section{Bohr's Complementarity Principle, Wiener's quantization of
thermodynamics, and a jump of critical exponents}

The equilibrium thermodynamics must follows from the nonequilibrium thermodynamics
in the limit as the viscosity tends to zero.
In this chapter, we first introduce the notion of viscosity $\nu$
and then let it tend to zero.
We introduce the viscosity in ``almost'' the same way as in hydrodynamics,
treating such an approach as the Wiener quantization.
In this chapter, we actually consider the limit as $\nu \to 0$.

Balescu wrote: ``The Maxwell construction does not provide a molecular
explanation of the phase transition below $T_{\mathrm c}$: it is merely an {\it
ad hoc\/} trick that works, provided we accept {\it a priori\/} the existence
of a coexistence region.''~\cite[p.~305]{20}.

Meanwhile, the problem involving the so-called Maxwell rule for the transition
``gas--liquid,'' which is a natural complement of the new concept (of
phenomenological thermodynamics) constructed above, is solved, as was described
in detail in~\cite{16}, by using the tunnel (or Wiener) quantization introduced
by the author already in his 1994 works; see~\cite{21, 22} and
also~\cite{23}--\cite{25}. We repeat here this quantization at a heuristic
level.

We can say that the quantization of thermodynamics is simply called for. We
have already mentioned Bohr's Complementarity Principle in the previous
section. Indeed, we have the phase space\footnote{``A symplectic
structure''.}\; in which the momenta are the extensive quantities~$V$ and~$-S$,
and the corresponding coordinates are~$P$ and~$T$. The usual quantization is of
the form
\begin{equation}
\widehat V=ih\frac\partial{\partial P}, \qquad -\widehat
S=ih\frac\partial{\partial T}. 
\end{equation}

Just as in~\cite{26}, let us invoke an analogy between the Schr\"odinger
equation and the heat equation.

A. {\sl The Schr\"odinger equation\/} corresponding to a noninteracting
particle without an external field is
\begin{equation}
-ih\frac{\partial\psi}{\partial t}=(ih\nabla)^2\psi. 
\end{equation}

The change of variables
$$
\psi=e^{\frac ih\mathbb S}
$$
leads to the equation
$$
\frac{\partial\mathbb S}{\partial t}+(\nabla\mathbb S)^2+ih\Delta\mathbb S=0.
$$
In this case, the quantization of the classical Hamilton--Jacobi equation
consists in the addition of the term $ih\Delta\mathbb S$.

B. {\sl The heat equation is}
\begin{equation}
-\nu\frac{\partial u}{\partial t}=(\nu\nabla)^2u, 
\end{equation}
where $\nu$ stands for the kinematic viscosity. The change of variables
$$
u=e^{-\frac{\mathbb S}\nu}
$$
leads to the equation
\begin{equation}
\frac{\partial\mathbb S}{\partial t}+(\nabla\mathbb S)^2+\nu\Delta\mathbb S=0. 
\end{equation}
The derivatives of this equation with respect to the coordinates are called the
{\it Burgers equations}. In this case, the {\it Wiener quantization\/} consists
in the addition of the viscous term.

\begin{remark}
In this special case, the Wiener quantization coincides with
the Euclidean quantization well known in field theory. In the general case,
this quantization corresponds to the passage from the Feynman path integral to
the Wiener path integral and is in essence presented in detail for physicists
in the book~\cite{27} by Feynman and Hibbs.
\end{remark}

In the Burgers equation, for $p=\partial\mathbb S/\partial x$, a shock wave
occurs as $\nu\to0$, i.e., a discontinuity of the $\theta$-function type,
whereas, in thermodynamics, we have a jump of the $\theta$-function type for
the transition ``gas--liquid.'' For the Burgers equation, the rule of ``equal
areas'' arises. For the ``gas--liquid'' transition, the Maxwell rule of equal
areas arises. In the heat equation, the tunnel quantization of energy is given
by the Heaviside operator $D_t=\frac\partial{\partial t}$ multiplied by the
viscosity, $\widehat D_t=\nu\frac\partial{\partial t}$. In thermodynamics, the
thermodynamical potential, the Gibbs energy, is equal to $\mu N$, where $\mu$
stands for the chemical potential and $N$ for the conjugate extensive quantity,
the number of particles. Hence, $\widehat N=\nu\frac\partial{\partial\mu}$, and
the role of time\footnote{Cf.~the Matsubara Green function, where the role of
imaginary time is played by the parameter $\beta=1/T$~\cite{28}.}\; is played
by $\ln-\mu$, because, under this quantization, the operator
$\nu\mu(\partial/{\partial\mu})$ corresponds to the Gibbs energy.

For us, the one-dimensional case $p_1=V$ and $q_1=P$ is of importance. In the
general case, the focal point (the point of inflection~\cite{29,30}) is of the
form $q\sim p^3$, i.e.,
\begin{equation}
P_{\mathrm{cr}}\sim(V-V_{\mathrm{cr}})^3=
\bigg(\frac{\rho_{\mathrm{cr}}-\rho}{\rho\rho_{\mathrm{cr}}}R\bigg)^3, 
\end{equation}
which corresponds to the classical critical index (the exponent) equal to
three. The asymptotic solution (as $\nu\to0$) of~\thetag{19} corresponding to
this point is expressed by the Weber function.

Therefore,
\begin{equation}
u(x)=\frac1{\sqrt\nu}\int_0^\infty e^{-\frac{px+\widetilde S(p)}\nu}\,dp,
\qquad \lim_{p\to0}\frac{\widetilde S(p)}{p^4}<\infty, \qquad \widetilde
S^{(4)}(p)|_{p=0}\ne0. 
\end{equation}

The solution $p_\nu(x)$ of the Burgers equation can be evaluated by the formula
\begin{equation}
p_\nu(x)=\nu\frac{\partial\ln u(x)}{\partial
x}=\frac{\int_0^\infty\exp\{\frac{-x\xi+\widetilde
S(\xi)}\nu\}\xi\,d\xi}{\int_0^\infty\exp\{\frac{-x\xi+\widetilde
S(\xi)}\nu\}d\xi}. 
\end{equation}
After the substitution  $$\frac\xi{\root4\of\nu}=y,$$ as $x\to0$ we obtain
\begin{equation}
p_\nu(x)\to_{x\to0}{\root4\of\nu}\cdot \text{const}. 
\end{equation}
In our case, the momentum $p_\nu(x)$ is the volume~$V$.

If the solution of the relation
\begin{equation}
x=\frac{\partial\widetilde S}{\partial p} 
\end{equation}
is nondegenerate, i.e., $$\frac{\partial^2\widetilde S}{\partial p^2}\ne0$$ at
the point $$\frac{\partial\widetilde S}{\partial p}=x,$$ then, in this case,
the reduced integral~\thetag{22} is bounded as $\nu\to0$. For this integral to
have a singularity of order $\nu^{1/4}$, we must apply to this integral the
fractional derivative $D^{-1/4}$ with respect to $x$. The value of $D^{-1/4}$
at the function equal to one, $D^{-1/4}1$, gives approximately $x^{1/4}$.

In our case, the pressure~$P$ plays the role of $x$, and the volume $V$ plays
the role of momentum $p$. Therefore, $V\sim P^{1/4},$ i.e.,
\begin{equation}
P_{\mathrm{cr}}\sim(V-V_{\mathrm{cr}})^4\sim
\Big(\frac{\rho-\rho_{\mathrm{cr}}}{\rho\rho_{\mathrm{cr}}}R\Big)^4. 
\end{equation}

Following Green~\cite{32}, D.~Yu.~Ivanov, a deep experimenter, poses the
following question: Why the deviations from the classical theory in the
critical opalescence are observed within the limits of hundredths of a degree
from the critical points, whereas the deviations in thermodynamical properties
show~\cite{33} a nonclassical behavior at a much larger distance from the
critical point? Professor Ivanov claims that rather many questions of this kind
have accumulated (see, for example,~\cite{34}), and all these questions mainly
deal with the behavior of practical systems. The point is that, from the point
of view of the developed theory of critical indices~\cite{35}, there must be a
drastic passage to the classical indices outside a neighborhood of the critical
point.

To make Ivanov's question understandable for persons who are not experts in
critical points, we paraphrase the question for the case of geometric optics,
when the sun rays are collected by a magnifying glass to a focus. If we were
created a special construction for the vicinity of the focus in which the paper
smoulders, then the experimenter could ask why the experiment gives a smooth
picture of transition in the double logarithmic coordinates and the indices are
preserved far away from the smouldering small vicinity of the focus. In the
present case, the smouldering paper can be compared with the small area of
opalescence (drastic fluctuations near the critical point) for which a separate
theory was constructed. At the same time, the special function defining the
point in wave theory (like the Weber function) can be continued quite smoothly
to a much wider domain in which the paper does not smoulder. In the opinion of
Ivanov, this fact is much more important than the fact that Wilson's theory
gives the index 4.82 rather than 4.3, whereas the latter is given by modern
experiments.

Can the experimental index 4.3 be explained in principle in the framework of
the conception presented by the author?

Whereas, in classical mechanics, there is no dependence on the Planck constant
$\hbar$, in classical thermodynamics we face a slow dependence of the viscosity
on $T$ and $\rho$, and thus {\it vice versa\/} as well. In our picture, the
``stretching'' of $P_{\mathrm{cr}}$ and $T_{\mathrm{cr}}$ in the experiment for
real gases (Figs.~10 and~11) is greater than in the van der Waals model, which
enables us to introduce the parameter $\nu^\varepsilon$ for the ``stretching''
(i.e., $P=\nu^\varepsilon V^3$) and obtain the index $\delta=4.3$ for
$\varepsilon=0.07$. Thus, in principle, the answer can be ``yes.''

It follows from what was said above that new critical indices arise only due to
quantization of the conjugate pairs $\{P,V\}$ and $\{T,S\}$. Thus, the
relationships between intensive quantities can be taken classical, because
everything is carried out under the assumption of infinitely small viscosity.
In this case, one can also pass to other coordinates, to the pressure and
density. Write, as usual,
$$
p=\frac{P-P_{\mathrm{cr}}}{P_{\mathrm{cr}}}, \qquad
\theta=\frac{T-T_{\mathrm{cr}}}{T_{\mathrm{cr}}}, \qquad
v=\frac{\rho-\rho_{\mathrm{cr}}}{\rho_{\mathrm{cr}}}.
$$
In the classical case, we have (\cite[p.~344]{20})
$$
p\sim v^3, \qquad v\sim\theta^{1/2}, \qquad \theta\sim v^2.
$$
In the classical case,
$$
p\sim\theta^{3/2}.
$$
In the tunnel quantum case, we obtain $\beta=0.375$ (cf.~\cite[p.~356]{20})
in the limit as $\nu\to0$, and thus not precisely (the stretching is not taken
into account). This is obtained for the van der Waals model quantized in the
tunnel way.

If $\nu>0$, then an uncertainty principle arises. Let us proceed to the
consideration of this principle.

Thus, for the Wiener quantization, one takes for the main operators the
Heaviside operator $\widehat D=\nu\frac\partial{\partial x}$ and the operator
of multiplication by $x$ rather than the momentum operator
$ih\frac\partial{\partial x}$ and the operator of multiplication by~$x$. We
have already defined the constant $\nu$ in~\cite{36} as viscosity. The tunnel
quantization mainly differs from the Euclidean and the ordinary ones in that it
is considered up to $O(\nu^k)$, where $k$ is an arbitrarily chosen number; this
means that this quantization is factorized with respect to $O(\nu^k)$. However,
first of all, one must define the space on which these operators act.

As is well known, the Heaviside operator is related to the two-sided Laplace
transform. This was shown already by van der Pol and Bremmer in~\cite{37}.
Introduce a family of functions $\varphi(p)$ to which we shall apply the
two-sided Laplace transform, namely,
\begin{equation}
\varphi(p)=\int_0^\infty e^{-p^2\xi}\Xi(\xi)\,d\xi, 
\end{equation}
and thus these functions by themselves are one-sided Laplace transforms of some
function,
$$
\widehat F_\lambda\Xi(\xi)=\int_0^\infty e^{-\lambda\xi}\Xi(\xi)\,d\xi
$$
for $\lambda=p^2$. Denote by $\widehat F_\lambda^\pm$ the two-sided Laplace
transform,
$$
\widehat F_\lambda^\pm\varphi(p)=\int_{-\infty}^\infty e^{-\lambda
p}\varphi(p)\,dp.
$$

If the functions $\Xi(\xi)$ are compactly supported and infinitely
differentiable, then the closure of the operator~$\widehat D=\nu D$ with
respect to this domain can be carried out in a Bergman space. Then the
functions $\Psi(x)=\widehat F_x^\pm\varphi(p)$ here become analogous to the
$\Psi$-functions in the Schr\"odinger quantization. Moreover,
$\Psi^*(x)=\Psi(x)$, because these functions are real-valued.

Let us note first of all that the squared function or, equivalently, the
squared dispersion $\Delta\widehat f$ of the operator $\widehat f$ is
$$
(\Delta\widehat f\;)^2=\big|(\widehat f-\overline f)^2\big|,
$$
and, since $\widehat f$ is not self-adjoint, the function $(\widehat
f-\overline f)^2$ need not be positive, and hence one must pass to its absolute
value. Therefore, the corresponding theorem for generic operators fails to hold
in general. However, for the operators $\widehat D=\nu D$ and $x$ on a reduced
function space, we obtain
$$
|\Delta\widehat D||\Delta x|\ge\frac\nu2.
$$

It can readily be seen that Weyl's proof (which is presented in the comments to
Sec.~16 of Chap.~II in~\cite{31}) can easily be transferred to the operators
$\widehat D$ and $x$ in the above function space.

Let us repeat the manipulations presented in~\cite{31} with regard to the fact
that $\widehat D$ satisfies the relation $\int\varphi\widehat
D\varphi\,dx=\frac12\int\widehat D\varphi^2\,dx=0$ on the class of functions in
question and $\int x\varphi^2\,dx=0$.

Consider the obvious inequality
\begin{equation}
\int_{-\infty}^{+\infty}\bigg|ax\psi+\frac{d\psi}{dx}\bigg|^2\,dx\ge0, 
\end{equation}
where $a$ stands for an arbitrary real constant. When evaluating this integral,
we see that
\begin{align}
\int x^2|\psi|^2\,dx&=\overline{(\Delta x)^2},\nonumber\\
\int\bigg(x\frac{d\psi^*}{dx}\psi+x\psi^*\frac{d\psi}{dx}\bigg)\,dx&=
\int x\frac{d|\psi|^2}{dx}\,dx=-\int|\psi|^2\,dx=-1,\nonumber\\
\int\frac{d\psi^*}{dx}\frac{d\psi}{dx}\,dx
&= -\int\psi^*\frac{d^2\psi}{dx^2}\,dx= \frac1{\nu^2}\int\psi^*|\widehat
D|^2\psi\,dx= \frac1{\nu^2}\overline{|\Delta\widehat D|^2}.
\end{align} 
We obtain
\begin{equation}
a^2\overline{(\Delta x)^2}-a+\frac1{\nu^2}\overline{|\Delta\widehat D|^2}\ge0. 
\end{equation}
For this quadratic trinomial (in a) to be positive for any value of a, the
condition
$$
4\overline{(\Delta x)^2}\frac1{\nu^2}\overline{|\Delta\widehat D|^2}\ge1
$$
must be satisfied, or
\begin{equation}
\sqrt{\overline{(\Delta x)^2}\ \overline{|\Delta\widehat D|^2}}\ge\frac\nu2. 
\end{equation}
Thus, the tunnel quantization explains both $\mu=0$  for photons and $\mu\le0$
for bosons.

Let us note some consequences of tunnel quantization for the ``quantum'' Bose
gas.

A specific feature of the photon gas, which is mentioned in~\cite[Secs.~62,~63]{11},
is that the number of particles in this gas, $N$, is a variable quantity
(rather than a given constant, which is the case for an ordinary gas).

Thus, since the number of particles $N$ in thermodynamics is conjugate to the
chemical potential, it follows that, if the number of particles is undefined,
then the chemical potential can be given precisely, $\mu=0$, under the
assumption that $\mu$ and $N$ are tunnel quantized and the uncertainty
principle holds.

A contradiction between the conception of the author and the conception of
physicists going back to Einstein is also removed. In the case of a gas for
which $N$ is fixed, we have
\begin{equation}
\sum^\infty_{i=0}N_i=N, 
\end{equation}
according to the relation in~\cite{11}, and the chemical potential $\mu$ can be
a small positive quantity. This is obvious, because $N_i\le N$; however, this
contradicts Einstein's original conception claiming that $\mu\le0$. This
contradiction is removed if the relationship of the uncertainty principle holds
for $\mu$ and $N$, because, if $\mu=0$, then $N$ can take infinite values as
well, and therefore the case $\mu>\nobreak0$ is impossible.

Thus, it can be said that both the scaling hypothesis and the hypothesis of
Wiener quantization do not agree in the vicinity of critical point with the old
thermodynamical conception of four potentials. However, the hypothesis of
Wiener quantization does not contradict the conception of four potentials,
namely, the hypothesis complements the conception, and this works not only near
the critical point but also on the entire domain ``gas--liquid'' by agreeing
with the Maxwell rule and by removing logical discrepancies in the Bose gas
theory.

We have explained the Wiener quantization, which enabled us to settle some
problems. Let us now also explain the second quantization for the Wiener
quantization. The principal element in Fock's approach is the
indistinguishability of particles. In our theory, this indistinguishability
follows from the main original axiom. Although there are no natural Hilbert
spaces here, in contrast to quantum mechanics, we can still obtain correct
distinguished representations and limits as $h\to0$
(see~\cite[Chap.~1, Appendix~1.A]{38}) and then,
in view of the new principle of indistinguishability
for grains, perform the second quantization of classical theory by introducing
the creation and annihilation operators. Certainly, this is possible only under
the condition of another ``identity principle'' than that used in~\cite{31},
namely, from the {\it principle of indistinguishability of particles in our
measurements}, which follows from the existence of macro-measuring instrument.

In classical mechanics, operators of this kind were introduced in~\cite{39, 40}
on the basis of the Sch\"onberg concept (see~\cite{41, 42}).\footnote{Our
considerations below are related to thermodynamics in nano-pores which was
described in detail in~\cite{48--50} and in~\cite{57}, and can be omitted for
the first reading.}

{{ Thus, the contemporary derivation of the Vlasov equation is obtained by
applying the method of second quantization for classical particles~\cite{38}.
In this case, as $N\to\infty$, one obtains a system for which the creation and
annihilation operators asymptotically commute,
\begin{align}
\dot u(p,q,t)= &\biggl(\frac{\partial U}{\partial q}
\frac\partial{\partial p}-p\frac\partial{\partial q}\biggr)u(p,q,t)
\nonumber\\
&+\int dp'dq'v(p',q',t)\biggl(\frac{\partial V(q,q')}{\partial q}
\frac\partial{\partial p}+
\frac{\partial V(q,q')}{\partial q'}\frac\partial{\partial p'}\biggr)u(p',q',t)u(p,q,t),
\\
\dot v(p,q,t)=& \biggl(\frac{\partial U}{\partial q}\frac\partial{\partial p}
-p\frac\partial{\partial q}\biggr)v(p,q,t)
\nonumber\\
&+\int dp'dq'u(p',q',t)\biggl(\frac{\partial V(q,q')}{\partial
q}\frac\partial{\partial p}+\frac{\partial V(q,q')}{\partial
q'}\frac\partial{\partial p'}\biggr)v(p',q',t)v(p,q,t),
\nonumber
\end{align} 
where $U(q_i)$ stands for an external field and $V(q_i,q_j)$ for the pairwise
interaction.

If one replaces $u$ and $v$ by the operators of creation and annihilation
$\widehat{u}$ and $\widehat{v}$ in the Fock space, then, after this change,
system~\thetag{32} becomes equivalent to the $N$-particle problem for the
Newton system.

However, according to the rigorous mathematical proof, this can happen only for
the case in which the classical particles are indistinguishable (from the point
of view of the notion of pile).

Only in this case does the projection from the Fock space to the
$3N$-dimensional space of $N$ particles give {\it precisely\/} the system of
Newton equations.

Note that the substitution
\begin{equation}
u(p,q,t)=\sqrt{\rho(p,q,t)}e^{i\pi(p,q,t)}, \quad
v(p,q,t)=\sqrt{\rho(p,q,t)}e^{-i\pi(p,q,t)} 
\end{equation}
reduces system \thetag{32} to the form
\begin{align}
&\dot\rho(p,q,t)=
\biggl(\frac{\partial W^t}{\partial q}\frac\partial{\partial p}
-p\frac\partial{\partial q}\biggr)\rho(p,q,t),
\\
&\dot\pi(p,q,t)=\biggl(\frac{\partial W^t}{\partial q}\frac\partial{\partial
p}-p\frac\partial{\partial q}\biggr)\pi(p,q,t)+\int dp'dq'\frac{\partial
V(q,q')}{\partial q'}\frac{\partial\pi(p',q',t)}{\partial p'}\rho(p',q',t);
\nonumber
\end{align} 
where $W^t(q)=U(q)+\int dq'V(q,q')\rho(p',q',t)dp'dq'.$

The first equation of system~\thetag{34} is the {{\it Vlasov}} equation
(see~\cite{44}), where $\rho$ stands for the distribution function and $W^t(q)$
for the dressed potential (see Sec.~2, the formulas beginning
with~\thetag{46}); the other equation is linear, and its meaning is discussed
in~\cite{45}. }}

\medskip

Note further a Wiener-quantum jump of the index at the points of the spinodal
of the liquid phase. The classical index of the spinodal is equal to 2, namely,
$P\sim V^2$, similarly to turning points in  quantum mechanics. The Airy
function corresponds to it. Similarly to~\thetag{21}--\thetag{23}, we obtain
\begin{equation}
\Psi(x)=\frac1{\sqrt\nu}\int_0^\infty e^{-\frac{px+\widetilde S(p)}\nu}\,dp,
\qquad \lim_{p\to0}\frac{\widetilde S(p)}{p^3}<\infty, \qquad \widetilde
S^{(3)}(p)|_{p=0}\ne0. 
\end{equation}

The solution $p_\nu(x)$ of the Burgers equation can be evaluated by the
formula~\thetag{22}. As $x\to 0$, after the change ${\xi}/{\root3\of{\nu}}=y$,
we obtain
\begin{equation}
p_\nu(x)\to_{x\to0}{\root3\of\nu}\times \text{const}. 
\end{equation}
In our case, the momentum $p_\nu(x)$ is the volume~$V$. Hence, similarly to the
consideration~\thetag{24--25}, we obtain $P\sim V^3$, and the index at the
points of the spinodal becomes equal to three.

\begin{remark} It is possible that an experimenter, when considering the
approaching of the critical isotherm for $T>T_{\mathrm{cr}}$ to the critical
point $\mu =0$, moves (due to the indeterminacy principle) towards increasing
values of~$N$, and hence towards increasing density, and arrives at the
spinodal of the liquid phase. This effect is similar to the accumulation of the
wave crest which overturns afterwards (a part of the particles outruns the
point of creation of the shock wave). In this case, the critical index 4.3
passes to the index 3 of the spinodal (and this index occasionally coincides
with the classical index of the critical point). This passage, which is
described by the Vlasov equation, was experimentally noticed in~\cite{34} and
in other works. Therefore, the experiments of Ivanov~\cite{34} and
Wagner~(\cite{46, 47}), where the modifications of the critical index $\delta$
from 4.3 to 3 were obtained when approaching the critical point, do not
contradict our conception.
\end{remark}

{{

For the creation of dimers, the author of the present paper used the creation
and annihilation operators for pairs of particles~(\cite{25} and~\cite{48--50})
and referred to this invention as the ultrasecond quantization. Experimenters
do not distinguish between dimers either, counting only their number (for
example, as was shown by Calo~\cite{51}, the presence of 5--7\% dimers leads to
the appearance of a cluster cascade).

Thus, we discover new relations, namely, an extension of the program ``partitio
numerorum'' in number theory from the point of view of the notion of Hartley
entropy, and indicate possible generalizations of quantization, which lead to
an extension of the Heisenberg indeterminacy principle~\cite{16}.

The ultrasecond quantization led to thermodynamics in nanocapillaries and
enabled one to obtain the superfluidity of liquids in nanotubes~\cite{48--50},
which was confirmed in experiments (see~\cite{52--54}).

}}

\medskip

A relationship between the parameters $\delta$ and $\beta$ follows from the
``classical'' thermodynamics. The relation for the compressibility index,
$$
\gamma=\beta(\delta-1),
$$
does not need the scaling hypothesis either. The corresponding inequality uses
convexity, which is closely related to tropical mathematics, which is the limit
as the viscosity tends to zero. The inequality becomes an equality as the
chemical potential tends to zero, $\mu\to0$, according to tropical
geometry~\cite{55}.

\section{Zeno line and relations for imperfect gas}

Experiments showed that the orthometric curve (the Zeno line) $Z=1$ ($PV=NT$)
on the $\{P,T\}$ plane is a line segment, and hence is completely determined by
the two endpoints of the segment, the points $T_B$ and $\rho_B$, where $T_B$ is
the well-known Boyle temperature and $\rho_B$ stands for the Boyle density,
which cannot be found experimentally, because the point $T=0$ is inaccessible.
This density is defined by extrapolation (see~Fig.~3). Only for water, the
straight line is somewhat bent in a domain near $T_B$, see~Fig.~4.

\begin{figure}[h]
\begin{center}
\includegraphics{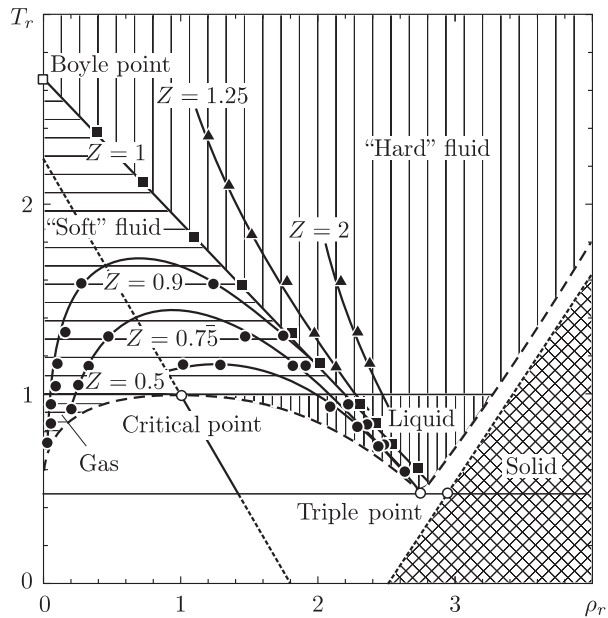}
\end{center}
\caption{$T$--$\rho$ diagram for gases corresponding to
simple liquids; $T_r=T/T_{ \text{cr}}$ and $\rho_r=\rho/\rho_{ \text{cr}}$.
The line $Z=\frac{PV}{NT}=1.0$ (Zeno line) on the phase diagram.
For states with $Z>1.0$ (hard fluids), the repulsive forces dominate.
For states with $Z<1.0$ (soft fluids), the attractive forces dominate.
The dotted line passing to the critical point and ending at the triple point
is the binodal (cf.~Fig.~8).}
\end{figure}

\begin{figure}[h]
\begin{center}
\includegraphics{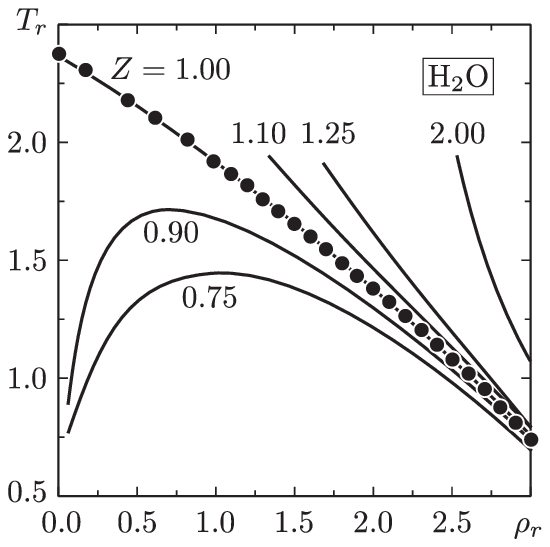}
\end{center}
\caption{$T$--$\rho$ diagram for water.}
\end{figure}

The Zeno line is precisely a straight line in the van der Waals model. Using
only heuristic considerations for the Lennard--Jones interaction potential, one
can show that this curve is almost a straight line indeed. These considerations
are of purely physical nature and use the existence of the so-called thermal
attractive potential.

Although the considerations presented here are not mathematically rigorous,
they elucidate some physical phenomena anew. The Boyle temperature $T_B$ and
some other quantities usually defined by using the van der Waals model are
treated here in a new way. Despite the fat that the logical reasoning is not
rigorous, it still perfectly agrees with the above rigorous conception. One can
prove in a mathematically rigorous way that the existence of a Zeno line (as a
segment of a straight line) can be regarded for pure gases as an additional
axiom, which can only be supported by heuristic considerations.

\subsection{Heuristic Considerations. The Role of Small Viscosity}

Following Clausius, experts in molecular physics usually argue by proceeding
from the symmetry of the motion of a molecule averaged in all six directions.
In the scattering problem, we use the principle of symmetry in all directions,
which is standard in molecular physics, but apply it to define not the mean
free path, but other molecular physics quantities. Therefore, the fraction of
all particles that moves head-on is 1/12. There are three such directions;
hence, one quarter of all molecules collide. \footnote{The arguments put
forward by Clausius concerning symmetry applied by Clausius to evaluate the
free path length and repeated here by the author are quite approximate.
However, these arguments do not influence on the values of ratios of the form
$T_B/T_{\mathrm{cr}}$.}

For the interaction potential, we consider the Lennard--Jones potential
\begin{equation}
u(r)=4\varepsilon\Big(\frac{a^{12}}{r^{12}}-\frac{a^6}{r^6}\Big), 
\end{equation}
where $\varepsilon$ is the energy of the depth of the well and $a$ is the
effective radius.

In the absence of an external potential, the two-particle problem reduces to
the one-dimensional radial-symmetric one.

Recall this passage.

Consider the two-body problem for particles of the same mass. Let us pass to
the new variables,
$$
r=r_2-r_1, \qquad R=\frac{r_1+r_2}2.
$$

This gives
\begin{equation}
r_1=R-r, \qquad r_2=R+r. 
\end{equation}
In the new variables, the kinetic energy is
\begin{equation}
T=\frac m2\dot R^2+ \frac\mu2\dot r^2, 
\end{equation}
where $\mu=m/2$ is the reduced mass of the system. Obviously,
 $\dot r=\dot r_2-\dot r_1$ is the relative velocity. The first summand
in~\thetag{39} is the kinetic energy of the relative motion of a
``$\mu$-point.'' The angular momentum of the system is
\begin{equation}
M=m[R\cdot\dot R]+\mu[r\cdot\dot r]. 
\end{equation}
In the new variables, the Lagrangian becomes
\begin{equation}
L(r,R,\dot r,\dot R)=\frac m2\dot R^2+\frac\mu2\dot r^2-U(r), 
\end{equation}
where $U(r)$ stands for the interaction potential.

Suppose that, as $t\to-\infty$, the velocities of the structureless particles
are equal to ${\bold v}_1^{\mathrm{in}}$ and
 ${\bold v}_2^{\mathrm{in}}$. This means that, as $t=-\infty$, the trajectories
of the particles approach straight lines. In terms of the variable $r=r_2-r_1$,
as $t\to-\infty$, the radius vector of the $\mu$-point asymptotically
approaches the function $r^{\mathrm{in}}=\rho+{\bold v}^{\mathrm{in}}t$, where
$\rho{\bold v}^{\mathrm{in}}=0$ and
 ${\bold v}^{\mathrm{in}}={\bold v}_2^{\mathrm{in}}-{\bold v}_1^{\mathrm{in}}$.

The constant vector $\rho$ is the impact parameter. The quantity $\rho$ is
equal to the distance between the straight lines along which the particles
would move in the absence of the interaction. After the collision, as
$t\to\infty$, the velocities of the particles are equal to
 ${\bold v}_1^{\mathrm{out}}$ and ${\bold v}_2^{\mathrm{out}}$. This means
 that~the radius vector $r(t)$ asymptotically approaches the function
 ${\bold r}^{\mathrm{out}}=c+{\bold v}^{\mathrm{out}}t$. The trajectories
 ${\bold r}^{\mathrm{in}}(t)$ and ${\bold r}^{\mathrm{out}}(t)$ are straight
lines. They are referred to as the incoming and outgoing asymptotes,
respectively. The value of the relative velocity in the in- and out-states in
preserved, namely, $|{\bold v}^{\mathrm{in}}|=|{\bold v}^{\mathrm{out}}|=v$.

The scattering process can be represented in the form of the transformation
\begin{equation}
{\bold v}^{\mathrm{in}}\to{\bold v}^{\mathrm{out}}={\bold n}v, 
\end{equation}
where $\bold n$ stands for the unit vector determining the kinematics of the
scattering.

According to the initial conditions,
$$
\bold M=\mu[{\bold r}\cdot{\bold v}], \qquad E=\frac\mu2v^2.
$$
Since ${\bold M}r=0$, it follows that the trajectory belongs to the plane of
the vectors $\bold r$ and $\bold v$. In the polar coordinates $r$, $\chi$, the
incoming asymptote corresponds to the value $\chi=0$.

\begin{figure}[t]
\begin{center}
\includegraphics{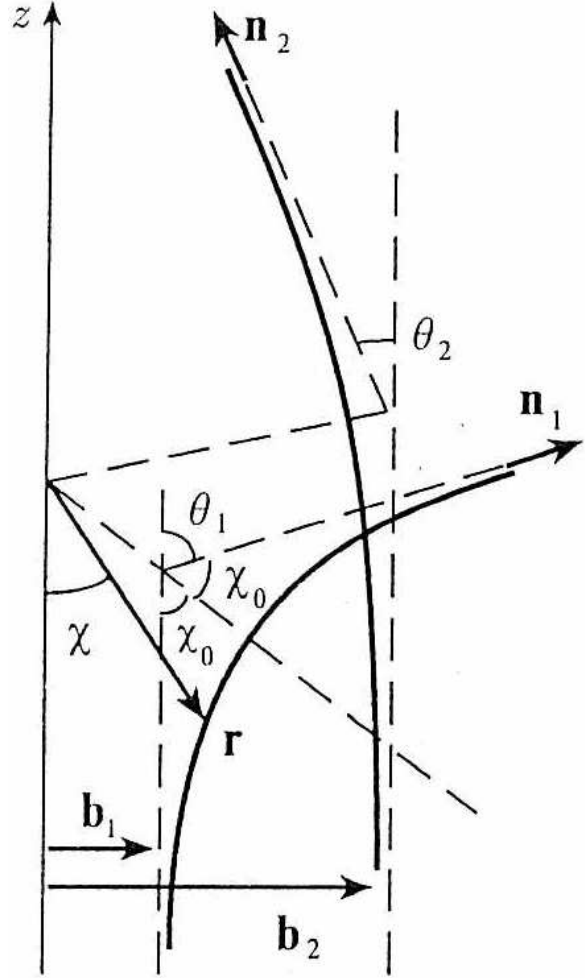}
\end{center}
\caption{Scattering for the Lennard--Jones potential.}
\end{figure}

The function $r(\chi)$ decreases, as $\chi$ increases, until it attains the
maximal value $r_{\max}=r_0$ at $\chi=\chi_0$, where the radial component of
the velocity vanishes. The outgoing asymptote corresponds to the value
$\chi=2\chi_0$. Both asymptotes are placed symmetrically with respect to the
line passing through the origin and the point of the trajectory which is the
nearest to the origin. In~dependence on the value of the impact parameter $b$,
the possible values of $\chi_0$ are in the interval $(0,2\pi)$. The observed
angle of scattering $\theta$, which is measures in the interval $(0,\pi)$ by
definition, is equal to $\theta=|\pi-2\chi_0|$. In Fig.~5, we show the incoming
and outgoing asymptotes in the case of Lennard--Jones potential energy. The
repulsion corresponds to the value $\rho_1$ and the attraction to the value
$\rho_2$~\cite{56}.

Two quantities are preserved in this problem, namely, the energy $E$ and the
momentum $M$. In the scattering problem, it is more convenient to consider
another constant (which is thus also preserved) instead of the momentum $M$,
namely, $b=\sqrt m\rho$, where $\rho$ stands for the impact parameter; thus,
\begin{equation}
M=\sqrt Eb. 
\end{equation}

Resolving the well-known relation
\begin{equation}
E=\frac{p^2}m+ \frac{M^2}{2mr^2}+u(r) 
\end{equation}
with respect to the energy~$E$, we obtain the attractive Hamiltonian $H$,
\begin{equation}
H=\frac{p^2/(2m)+u(r)}{1-b^2/r^2}, \qquad a<r\le b. 
\end{equation}
This simple transformation, if we take the influence of the small viscosity
into account, enables us to modify the standard scattering problem in such a
way that both the quantity $T_B$ and the quantity $T_{\mathrm c}$ obtain a new
meaning.

The phenomenon which we have described above by using the example of wells is
actually a continuous process (which is established for a given temperature) of
random creation of dimers and cleaving of dimers by quick monomers. We may
speak only of the percent of dimers at a given temperature.

The repulsive Hamiltonian is separated from $H$ by a barrier. Repulsive
particles make obstacles in the way of particles of the Hamiltonian $H$, by
creating a ``viscosity.''

As the temperature decreases, the height of the barrier grows up to the value
$E_{\mathrm{cr}}=0.286 \varepsilon$, and then starts reducing (see~Fig.~6).
According to rough energy estimates~ \cite{58}, for lesser temperatures, an
additional barrier must be formed as the clusters are created.

This barrier can be given for neutral gases and methane by germs of droplets,
i.e., three-dimensional clusters that contain at least one molecule surrounded
by other molecules (a prototype of a droplet).\footnote{By a ``barrier'' we
mean an obstacle to a collision of particles; a ``shell'' of surrounding
particles defends the given particle from an immediate blow. In mathematics, a
``domain'' is an open region containing at least one point.}

\begin{figure}[h]
\begin{center}
\includegraphics{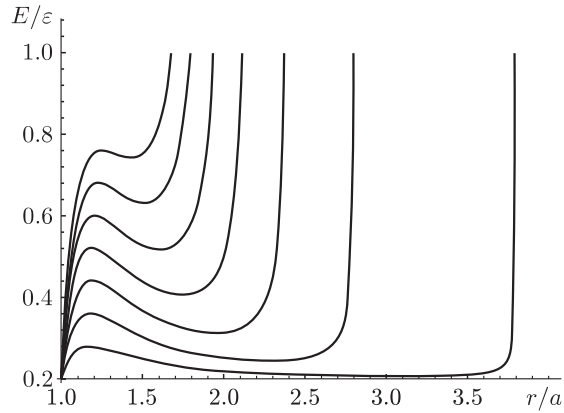}
\end{center}
\caption{The values of $E(b,r)$ for diverse values of the
impact parameter (from left to right): 1.81378, 1.89344, 2.00178, 2.1559,
2.39252, 2.80839, and 3.79391. The critical point is not shown; the minimum
disappears at $\rho=1.75441$.}
\end{figure}

However, to study the penetration through the barrier of the incident particle,
we must plot~$E$ along the $y$ axis and turn the wells upside down.  Then the
minimum becomes the barrier and the maximum becomes the depth of the well (see
Fig.~7).

\begin{figure}[h]
\begin{center}
\includegraphics[width=6truecm]{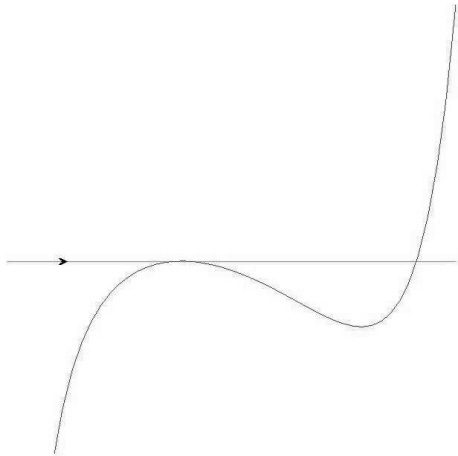}
\end{center}
\caption{The trap for a fictitious particle in the the
center of mass coordinate system. The radius vector $r$ of the $\mu$-point is
marked on the abscissa axis. The particle falls from the left from the point
$r=b$, where $b$ stands for the impact parameter. In the original
problem~\thetag{43} (before the change of variables $M=\sqrt Eb$), the relation
$r=b$ is attained as $r\to\infty$.}
\end{figure}

A dimer can be formed in a classical domain if the scattering pair has an
energy equal to the barrier height, slipping into the dip in ``infinite" time
and getting stuck in it as the result of viscosity (and hence of some small
energy loss), because this pair of particles, having lost energy, hits the
barrier on the return path. If the pair of particles has passed above this
point, then the viscosity may be insufficient for the pair to become stuck:
such a pair returns above the barrier after reflection. Therefore, only the
existence of a point~$E=E_{\max}$ plus an infinitesimal quantity, where
$E_{\max}$ is the upper barrier point, is a necessary condition for the pair to
be stuck inside the dip; $E_{\max}^{\mathrm{cr}}$ is the height of the maximum
barrier.

We can compare the values $T_{\mathrm{cr}}$ with the values
$E_{\max}^{\mathrm{cr}}$ in the table below.

\bigskip
\begin{center}
\begin{tabular}{|c|c|c|c|}
\hline
{Substance} & {$\varepsilon$, K} & {$T_{cr}/4$}  & {$E_{cr}\cdot\varepsilon/k$}\\
\hline
{$Ne$} & 36.3 & 11 & {10.5} \\
\hline
{$Ar$} & 119.3 & 37 & {35}\\
\hline
{$Kr$}& 171 & 52 & {50}\\
\hline
{$N_2$} & 95,9  & 31 & {28}\\
\hline
{$CH_4$} & 148.2 & 47 & {43}\\
\hline
\end{tabular}
\end{center}
\bigskip

Above the value $E_{ \text{B}}=0.8\varepsilon$, the trap disappears. At the
value $0.286\varepsilon$, the depth of the trap is maximal and corresponds to
$T_{\mathrm{cr}}=\frac{1.16\varepsilon}k$. For neon and krypton, as can be seen
from the table, the concurrence is sufficiently good. Because
$T_B=3.2\varepsilon/k$, it follows that $T_B/T_{\mathrm{cr}}=2.7$, which
corresponds to the known relation of ``the law of corresponding
states''~\cite{59}.

The temperature corresponding to $4E_{ \text{B}}/k$, is the temperature above
which dimers do not appear. Exactly this is what we call the Boyle temperature
(in contrast to~\cite{11}).

In fact, an application of the Clausius approach to pairwise interaction gives
a pairwise interaction with respect to the Lennard--Jones potential for two
Gibbs ensembles of noninteracting molecules. This leads to the presence of a
small friction for a single pair.

The difference $E_{\max}-E_{\min}$ is equal to the energy needed for a particle
lying at the bottom of the potential well to overcome the barrier. The value
$E_{\max}$ corresponds to the temperature given by $E_{\max}=RT$, where $R$
stands for the universal gas constant. According
to graph~6,
$E_{\min}$ corresponds to the energy $PV$. Therefore, $E_{\min}/E_{\max}\le1$
is the compressibility factor, $Z=PV/RT$. The temperature at the point
$E_{\min}=E_{\max}$ is equal to the Boyle temperature.

The dressed or ``thermal'' potential $\varphi(r)$ is attractive~\cite{60}. In
addition, because the volume~$V$ is a large parameter, it follows that, if the
quantity $\varphi(r)=({N\varepsilon a}/{\root3\of V})U\big(r/{\root3\of
V}\big),$ where $U\big(r/{\root3\of V}\big)$ is a smooth function and $N$
stands for the number of particles, is expanded in terms of $1/\root3\of V$,
then
\begin{equation}
U\Big(\frac r{\root3\of V}\Big)=C_1+ \frac{C_2r}{\root3\of
V}+\frac{C_3r^2}{(\root3\of V)^2} +O\bigg(\frac1{(\root3\of V)^3}\bigg). 
\end{equation}
Expanding
\begin{equation}
C_1+r^2=\frac{(r-r_0)^2}2+\frac{(r+r_0)^2}2, 
\end{equation}
where $C_1=r_0^2$, we can separate the variables in the two-particle problem as
above and obtain the scattering problem for pairs of particles and the problem
of their cooperative motion for $r_1+r_2$. The term $C_2r/\root3\of V$ does not
depend on this problem and the correction $({a\varepsilon}/{\root3\of
V})NO\big(1/V\big)$ is small.

Then, in the scattering problem, an attractive quadratic potential (inverted
parabola multiplied by the density or, to be more precise, by the
concentration, which we denote by the symbol~$\rho$ as well, because the target
parameter does not occur below) is added to the Lennard--Jones interaction
potential.

For this problem, we can find just as in~\thetag{44--45}, for all $\rho=N/V$, a
point corresponding to the temperature at which the well capturing the dimers
vanishes, and thus determine the so-called Zeno line. It is actually a straight
line (up to 2\%), on which $Z=E_{\min}/E_{\max}=1$ (i.e., an ideal curve).

Let us clarify this fact in more detail.

We can treat the repulsing potential as a potential creating a small viscosity.

Let us find the total energy of the attractive Hamiltonian,
$$
E=\bigg(\frac{mv^2}{2(1-b^2/r^2)}\bigg)+\frac{\Phi(r)}{1-b^2/r^2}\,, \qquad
\Phi(r)=u(r)-\rho r^2.
$$
The first term is negative for $r\le b$ and the other term is positive for
$b>r>a$ (i.e., the more is the speed, the less is energy). The mean speed is
temperature.

Let us make the change of variables
$$
\frac ra=r', \qquad \frac ba=\widetilde b,
$$
and get rid of~$a$. In what follows, we omit both the tilde and the prime.

For a given $b$, the minimum $r_1$ and the maximum $r_2$
(see the graph no.~1 in~\cite{61}) are defined by the relation
\begin{equation}
\frac{dE}{dr}=0. 
\end{equation}

This gives $E_{\max}$ and $E_{\min}$. These values coincide at some point
$b=b_0$, and hence
\begin{equation}
\frac{d^2E}{dr^2}=0 
\end{equation}
at the point $r_0$, i.e., $E_{\max}=E_{\min}$, and this is the very Zeno line.

Let us construct the curve $Z_{\min}=E_{\min}/E_{\max}$ minimal with respect to
the target parameter as a function of $\rho$. Let us find the point
$Z=E_{\min}/E_{\max}$ for $E_{\max}=E_{\max}^{\mathrm{cr}}$ and find the
corresponding point on the curve $Z_{\min}(\rho)$. This point is equal to $Z_{
\text{cr}}=0.29$, i.e., to the critical value of the compressibility factor $Z$
for argon.

\begin{figure}[h]
\begin{center}
\includegraphics{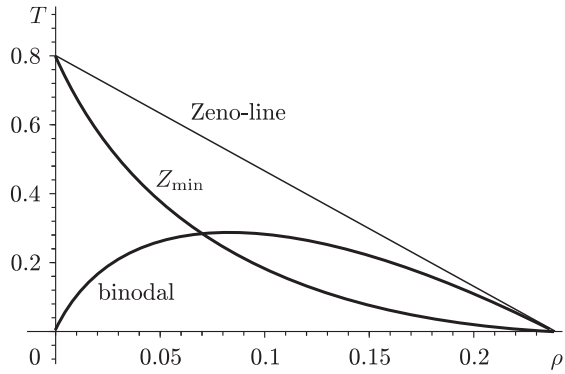}
\end{center}
\caption{The binodal, the Zeno line, and the curve $Z_{\min}$.
This heuristic binodal does not coincide with the experimental one,
whereas the Zeno line and $Z_{\mathrm{cr}}$ are close to the corresponding
experimental curves.}
\end{figure}

In order to obtain a binodal according to some ``heuristic principle,'' we must
subtract the curve $Z_{\min}(\rho)$~\cite{62} from the Zeno line. This gives
the graph shown in Fig.~8.

\subsection{Consideration of Interaction: Nonideal Gas}

At first glance, it looks as if the notion of new ideal gas replaces the famous
relation, which was approved for many years,
\begin{equation}
PV=NT 
\end{equation}
(which, moreover, served as an analogy for the main economical law, Irving
Fisher's formula; which is used to calculate the ``turnover rate'' of
capital~\cite{63}). This could be surprising indeed. However, this is not the
case. The relation $PV=NT$ or, equivalently, $PV=RT$ (because the number of
particles in the vessel remains the same) defines an imperfect gas and, in
contemporary experimental thermodynamical diagrams, it is called the {\it Zeno
line} or, sometimes, the ``ideal curve,'' the ``Bachinskii parabola,'' or the
orthometric curve.

On the diagram $(\rho,T)$ for pure gases, this is the straight line $Z=1$. This
line is a most important characteristic feature for a gas which is imperfect.
Since, for imperfect gases, it has been calculated experimentally and is an
``almost straight'' line on the $(\rho,T)$ diagram, it follows that the Zeno
line is determined by two points, $T_B$ and $\rho_B$, called the ``Boyle
temperature'' and the ``Boyle density.'' In contrast to $Z_{\mathrm{cr}}$,
these points are related to the interaction and scattering of a pair of gas
particles in accordance with the interaction potential specific for this gas,
as was shown in Section~1 and in other papers of the author (see,
e.g.,~\cite{64}). Therefore, the Zeno line on which the relations
\begin{equation}
PV=NT, \qquad \frac\rho{\rho_B}+\frac T{T_B}=1, \qquad \frac NV=\rho, 
\end{equation}
are satisfied, where $\rho$ is the density (the concentration), is a
consequence of pairwise interaction, and thus is a relation for an imperfect
gas.

The correction related to the existence of the Zeno line leads to a
differential equation~\cite{65} whose numerical solution yields an alteration
to the gas spinodal for every particular pure gas. For argon and $CO_2$, this
modification is shown in Fig.~9.\footnote{By heuristic considerations related
to the scattering problem (Section~1), the final point~\cite{66} of the gas
spinodal is equal to $Z=3/2\,Z_{\mathrm{cr}}$, and the spinodal can be
approximated by a line segment. (Ideally, at infinite time, a fictitious
particle (a pair) falls to the bottom due to the friction, i.e., the orbit of
this particle is circular, and thus one degree of freedom disappears. This
means that the compressibility factor $Z=0.444$ at the point
$E_{\max}^{\text{cr}}$ is reduced by the factor 2/3.). This makes it possible
to construct two points near $P=0$, $Z=Z_{\mathrm{cr}}$ by the theories of a
new ideal (Bose) gas and by the fact that the chemical potential of the gas is
equal to the chemical potential of an ideal liquid, and thus to approximately
reconstruct the Zeno line.}

The distribution of number theory, as opposed to the ``Bose--Einstein
distribution,'' does not contain the volume $V$. Let us consider the
distribution of number theory multiplied by unknown
function~$\varphi_{\gamma_0}(V)$ which does not vary for $\gamma\ge\gamma_0$
and $T\le T_{\mathrm{cr}}$. Then it follows from~\thetag{51} that
\begin{gather}
P=\frac{\varphi'_{\gamma_0}(V)T^{\gamma_0+2}}{\Gamma(\gamma_0+2)}
\int^\infty_0\frac{\varepsilon^{\gamma_0+1}\,d\varepsilon}
{e^{-\kappa}e^\varepsilon-1}, \qquad
\varphi'_{\gamma_0}=\frac{d\varphi_{\gamma_0}}{dV},
\\
\varphi_{\gamma_0}'(V)\operatorname{Li}_{\gamma_0+2}(y)=
\frac\rho{T_B^{\gamma_0+1}\big(1-\frac\rho{\rho_B}\big)^{\gamma_0+1}}, \quad
\rho=\frac RV, \quad y=e^\kappa, \quad \kappa=\frac\mu T.
\nonumber
\end{gather} 
The differential equation for $\varphi_{\gamma_0}$ is
\begin{equation}
\frac{V\varphi'_{\gamma_0}(V)}{\varphi_{\gamma_0}(V)}
\frac{\operatorname{Li}_{\gamma_0+2}(y)} {\operatorname{Li}_{\gamma_0+1}(y)}=1,
\qquad V=\frac R\rho. 
\end{equation}
See Fig.~9.

\begin{figure}[h]
\begin{center}
\includegraphics{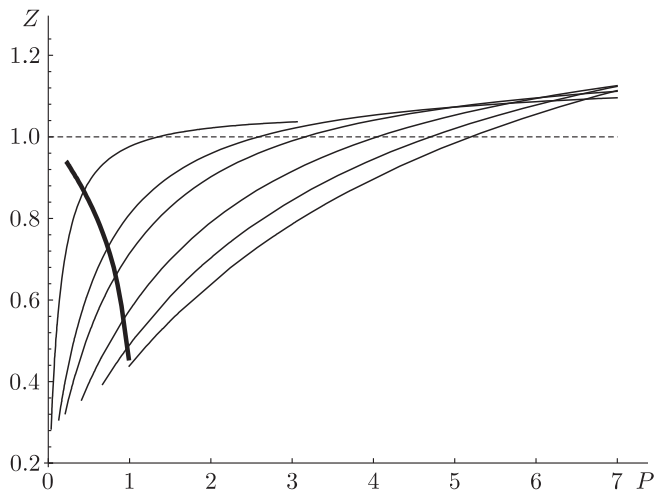}
\end{center}
\caption{The dotted line shows the Zeno line $Z=1$. The
bold line is the critical isotherm of an imperfect gas (argon) calculated
theoretically; the thin lines correspond to the isochores of an~imperfect gas
for $T<T_{\mathrm{cr}}$. Their initial points lie on the gas spinodal.}
\end{figure}

\begin{remark}
The notion of Lagrangian manifold, which is the ``equation of
state'' defining a two-dimensional surface in the four-dimensional phase space,
which was introduced by the author in~\cite{69} (see MSC2010 (Mathematics
Subject Classification 2010) http://www.ams.org/msc/, the section ``53D12
Lagrangian submanifolds; Maslov index'') enables one to carry out this
multiplication by the function $\varphi_{\gamma_0}(V/{V_0})$ without violating
the Lagrangian property, and thus the basic relationships for the free energy,
internal energy, and thermodynamical potential are preserved.
\end{remark}

\section{Ideal liquid}

Let us now pass to the notion of ideal liquid. For an expert in mathematical
physics, an ideal liquid is an incompressible liquid. In our mathematical
conception of thermodynamics, we shall abide by this definition. In this case,
on the Zeno-line on the plane $\{P,Z\}$, for $Z=1$, the point $P(T,\rho)$
in~\thetag{51} is defined uniquely. The isotherm $T= \text{const}$ is a
straight line. The second point is obviously the spinodal point.

As is well known, the passage from the gaseous state to the liquid one is
accompanied by an entropy drop. Naturally, the entropy, which determines the
measure of chaotic behavior, is less for the liquid state than for the gaseous
state. At the same time, the general property of ``choosing'' a subsystem with
the greatest chaoticity among all possible subsystems leads to the property of
constant entropy of the liquid, which was noted both theoretically and
experimentally, even if the temperature tends to the absolute zero~(\cite{67,
68}) (the entropy tends to $\ln 2$).

In our model of ideal liquid as an incompressible liquid, we suppose in
addition that the maximum of the entropy on a given isotherm (i.e., as $\mu \to
0$) does not vary when the temperature varies.

The big thermal potential is of the form
\begin{equation}
\Omega=-PV_\gamma=-\frac{\pi^{\gamma+1}V_\gamma T}{\Lambda^{2(1+\gamma)}}
\frac1{\Gamma(2+\gamma)}\int_0^\infty\frac{t^{1+\gamma}\,dt}{(e^t/z)-1}=
\frac{-\pi^{\gamma+1}V_\gamma T^{2+\gamma}}{\Lambda^{2(1+\gamma)}}
\operatorname{Li}_{2+\gamma}(z), \quad z=e^{\mu/T}, 
\end{equation}
where $\Lambda$ is a constant, its own for every substance (as a rule, it
depends on mass; however, we try to avoid mass by passing from density to
concentration).

According to~\cite{11}
the entropy is of the form
\begin{gather}
S=-\bigg(\frac{\partial\Omega}{\partial T}\bigg)_{V,\mu}=(2+\gamma)
\frac{T^{1+\gamma}}{\Lambda^{2(1+\gamma)}}\operatorname{Li}_{2+\gamma}(z)-
\frac{T^{1+\gamma}}{\Lambda^{2(1+\gamma)}}\operatorname{Li}_{1+\gamma}(z)
\frac\mu T
\nonumber\\
=\frac{\pi^{\gamma+1}T^{1+\gamma}}{\Lambda^{2(1+\gamma)}}
\Big[(2+\gamma)\operatorname{Li}_{2+\gamma}(z)-\operatorname{Li}_{1+\gamma}(z)\frac\mu
T\Big].
\end{gather} 
The maximum at $\mu=0$ is
\begin{equation}
S_{\mu=0}=\Big(\frac\pi{\Lambda^2}\Big)^{\gamma+1}(2+\gamma)\zeta(\gamma+2)T^{\gamma+1}. 
\end{equation}
We are interested in 
in the case  $\gamma<0$ as well.

Thus, we have two unknown constants, namely, $\Lambda$ and the value of the
entropy $S_{\mu=0}=\operatorname{const}$
\footnote{The exact value of~$\Lambda$ was calculated in~\cite{13},~\cite{14};
in~\cite{17} it was calculated
in a way that is simpler for physicists.}.
These two constants can be defined from
the experimental value of the critical point of the liquid phase at the
negative pressure (see Section~6 below), namely, from the minimum point of the
pressure for a given simple liquid of the value $\gamma$ at this point and from
the temperature. This point is absent in the van der Waals model. This point is
present in our model of liquid phase\footnote{The constant~$\Lambda$
can be calculated exactly, see the preceding footnote.}.

For example, for water, we obtain $S_{\mu=0}=3.495$ and $\Lambda=3.74$.
However, the computation is carried out under the assumption that the Zeno-line
is a line segment, whereas this segment becomes curvilinear for water at low
temperatures (see Fig.~4).

According to the van der Waals conception, we normalize as follows:
\begin{equation}
T^{\mathrm{red}}=\frac T{T_{\mathrm{cr}}}, \qquad P^{\mathrm{red}}=\frac
P{P_{\mathrm{cr}}}. 
\end{equation}

\begin{figure}[h]
\begin{center}
\includegraphics{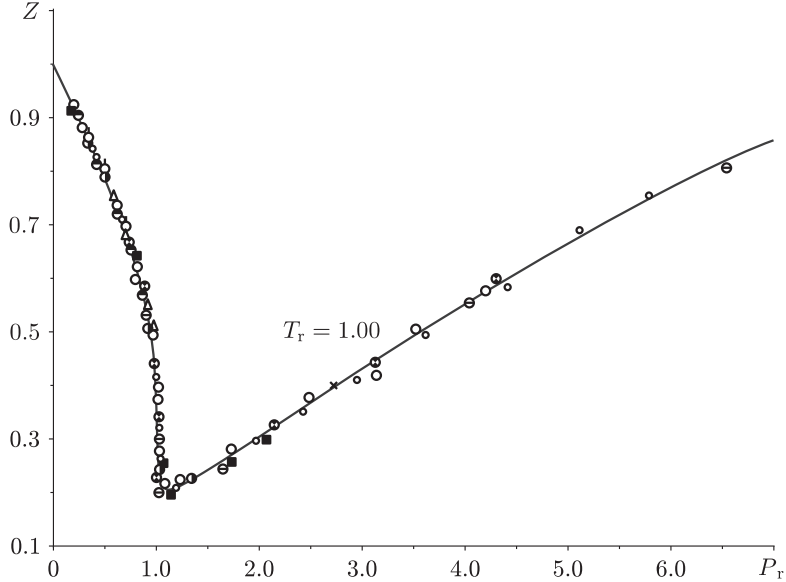}
\end{center}
\caption{Experimental graph for the different gases,
including those for methane, ethylene, ethane, propane, $n$-butane, isopentane,
$n$-heptane, nitrogen, carbon dioxide, and water. Each gas is equipped with a
particular symbol on the graph.}
\end{figure}

\begin{figure}[h]
\begin{center}
\includegraphics{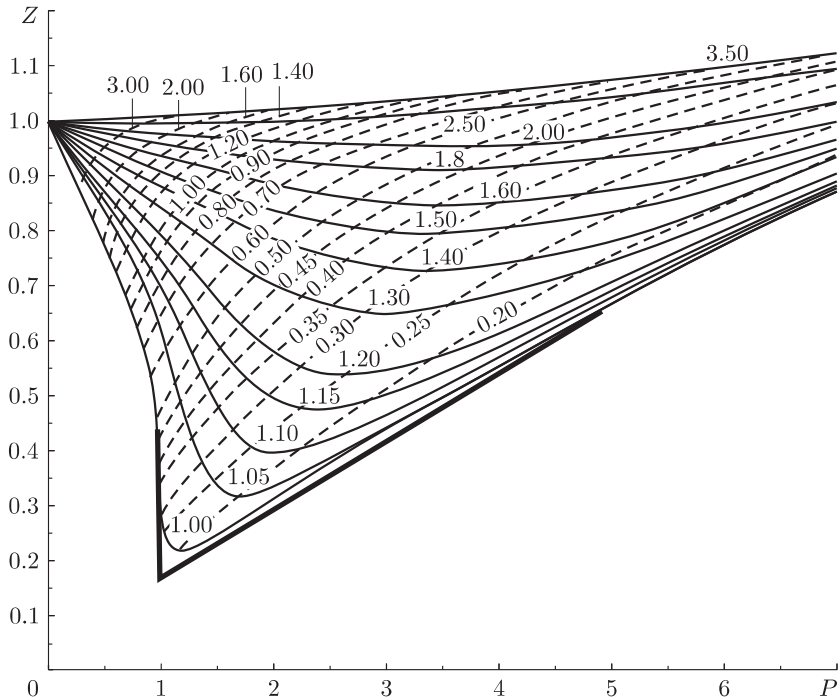}
\end{center}
\caption{The continuous lines are the experimental
isotherms for $T\ge T_{\mathrm{cr}}$ for methane, and the dotted lines are
experimental isochors. The theoretical critical isotherm coincides with the
experimental isotherm up to $Z=0.29$ and is continued by a straight line up to
the point $P=1$, $Z=1/\rho_0=0.14$ (see \thetag{58, 59}). Further, at an acute
angle, a tangent to the experimental isotherm at a point of Zeno-line is drawn.
The straight line from the point $P=1,Z=0.14$ to the point of tangency is the
critical isotherm of the ideal liquid phase. It can be seen by comparing the
figure with Fig.~10, the theoretical isotherm thus obtained corresponds to the
isotherm of the ``law of corresponding states'' for the gases indicated in
Fig.~10.}
\end{figure}

Denote by $\rho_0$ the solution of the equation
\begin{equation}
\frac1{T_B^{\mathrm{red}}}+\frac\rho{\rho_B}=1, 
\end{equation}
where $T_B$ stands for the Boyle temperature and
$T_B^{\text{red}}=\frac{T_B}{T_{\text{cr}}}$ is a dimensionless quantity,
$T^{\text{red}}=\frac T{T_{\text{cr}}}$. Then
\begin{equation}
\frac{\rho_0}{\rho_B}=1-\frac1{T_B^{\mathrm{red}}}, \qquad
\rho_B^{\mathrm{red}}=\frac{\rho_B}{\rho_0}. 
\end{equation}

Hence, the locus of the spinodal points\footnote{That is, of the endpoints of
the metastable state of the liquid phase.}\; is given by the formula
\begin{equation}
P^{\mathrm{red}}=
\frac1{\Big(1-\frac1{T_B^{\mathrm{red}}}\Big)}T^{\mathrm{red}}\Big(1
-\frac{T^{\mathrm{red}}}{T_B^{\mathrm{red}}}\Big)Z(\gamma). 
\end{equation}

Hence, for $\gamma>0$,
\begin{equation}
Z_{\mathrm{cr}}^{\gamma(T^{\mathrm{red}})}=
\frac{\zeta(2+\gamma(T^{\mathrm{red}}))}{\zeta(1+\gamma(T^{\mathrm{red}}))}. 
\end{equation}
Recall that $\gamma(T)$ can be calculated from the algebraic relation
$S_{\mu=0}=\mathrm{const}$.

If we use the Maxwell condition which states that the transition from gas to
liquid occurs for the same chemical potential, pressure, and temperature, then
we can construct the so-called {\it binodal}. The binodal thus constructed
coincides with the experimental one, in contrast to the van der Waals binodal
(Fig.~1\,(b)) and to the binodal presented in Fig.~8.

\section{Negative pressure and a new critical point of possible transition from
liquid to ``foam''}

First of all, using the Wiener quantization of thermodynamics, we shall now
clarify why in Section~4 we have derived the condition
$S|_{\mu=0}=\text{const}$, where the constant does not depend on the
temperature, from the model of incompressible fluid.

By the Bachinskii relation on the Zeno line, the model of incompressible fluid
leads to a rigid relationship between the density (the concentration)~$\rho$
and the temperature. Since the value~$N$ is undefined for $\mu=0$ (i.e., the
concentration $\rho$ is undefined), it follows that the temperature is
undefined as well and, by the indeterminacy principle, the entropy takes a
constant value and can be defined. (Note in addition that, for $\mu=0$, the
activity is equal to one for any (undefined) temperature.) This very fact means
that $S|_{\mu=0}=\text{const}$. As we shall see below, the value of this
constant is uniquely defined by the new critical point of transition from the
liquid to a ``foam'' (a disperse state,~\cite{58}). The value of the same point
also defines the constant $\Lambda$ in the definition of $\Omega$, and thus,
after computing $\varphi_\gamma(V/V_{\text{cr}})$, one can completely describe
the distribution function for the gaseous branch in thermodynamics. Note that
the indeterminacy at the spinodal point agrees with the experiments of the
Academician Skripov and his school on the absolute instability of the spinodal
point.

Let us now proceed with negative values of~$Z$.

As is known, the Bose--Einstein distribution is obtained as the sum of terms of
an infinitely decreasing geometric progression. If the progression is bounded
by the number $N$, then the potential becomes
\begin{equation}
\Omega=-T\sum_k\ln\bigg(\frac{1-\exp\frac{\mu-\varepsilon_k}TN}{1-
\exp\frac{\mu-\varepsilon_k}T}\bigg). 
\end{equation}

What is the relationship\footnote{Physicists which are not interested in
Euler--Maclaurin type estimates for passages from sums to integrals can omit
the below scheme of proving these estimates and proceed with
formulas~\thetag{68} and further.} \ between $E_i$ and $\gamma$?

(1) Ultrarelativistic case. Here $E=cp$.
\begin{equation}
E_{i+1}-E_i=\int_{E_i}^{E_{i+1}}cp\,p^2\,dp=
\frac14\bigg(p^4(E_{i+1})-p^4(E_i)\bigg)\sim
\frac34cp^3(E_i^{i+1})=\frac3{c^2}(E_i^{i+1})^3. 
\end{equation}

\medskip

(2) Nonrelativistic case. Here $E=\frac{p^2}{2m}$.
\begin{gather}
E_{i+1}-E_i=\int_{E_i}^{E_{i+1}}\frac{p^2}{2m}p^2\,dp=
\frac1{2m}\bigg(\frac{p^5}5(E_{i+1})-\frac{p^5}5(E_i)\bigg)\\
\frac1{2m}\bigg(\frac{(\sqrt{2mE_{i+1}})^5}5-\frac{(\sqrt{2mE_i})^5}5\bigg)\cong
\mathrm{const}(E_i^{i+1})^{3/2}.
\nonumber
\end{gather} 

\medskip

(3) Consideration of the degrees of freedom. $E=\frac{p^{2+\sigma}}{2mp_0}$.
Here
\begin{equation}
E_{i+1}-E_i=\int_{E_i}^{E_{i+1}}\frac{p^{2+\sigma}}{2m}p^2\,dp\cong
\mathrm{const}(E_i^{i+1})^{(4+\sigma)/(2+\sigma)}. 
\end{equation}

By~\thetag{13},
$$
\gamma=\frac{1-\sigma}{2+\sigma}
$$
and thus $\gamma <0$ for $\sigma >1$.

As was proved in~\cite{13--14}, passing to the limit in the Euler--Maclaurin
formula, we have proved that
\begin{equation}
N=\frac1{(\gamma+1)\Gamma(\gamma+1)}\int_0^\infty
\left\{\frac1{e^{b\xi}-1}-\frac N{e^{Nb\xi}-1} \right\}\,d\xi^\alpha. 
\end{equation}
In particular, for $\gamma=-1/2$,
\begin{equation}
N=\frac1{\Gamma(3/2)}\int_0^\infty \left\{\frac1{e^{b\xi^2}-1}-\frac
N{e^{Nb\xi^2}-1} \right\}\,d\xi. 
\end{equation}

The absolute value of the derivative of the integrand can readily be estimated
by using the~identities presented below. By the Euler--Maclaurin bounds, this
shows that one can pass from the sums of the form~\thetag{62} to the
corresponding integrals with the accuracy needed here.

Hence, writing $N_{\operatorname{cr}}=k_0$, we obtain the following formula for
the integral at $\mu=0$:
\begin{equation}
\mathcal E=
\frac1{\alpha\Gamma(\gamma+2)}\int\frac{\xi\,d\xi^\alpha}{e^{b\xi}-1}=
\frac1{b^{1+\alpha}}\int_0^\infty\frac{\eta d\eta^\alpha}{e^\eta-1}, 
\end{equation}
where $\alpha=\gamma+1$. Therefore,
\begin{equation}
b=\frac1{\mathcal
E^{1/(1+\alpha)}}\left(\frac1{\alpha\Gamma(\gamma+2)}\int_0^\infty\frac{\xi
\,d\xi^\alpha}{e^\xi-1}\right)^{1/(1+\alpha)}. 
\end{equation}

We obtain
\begin{align}
&\int_0^\infty
\left\{\frac1{e^{b\xi}-1}-\frac{k_0}{e^{k_0b\xi}-1}
\right\}\,d\xi^\alpha =\frac1{b^\alpha}\int_0^\infty\left(\frac1{e^\xi-1}
-\frac1\xi\right)\,d\xi^\alpha 
\\
&\quad+\frac1{b^\alpha}\int_0^\infty\left(\frac1\xi-
\frac1{\xi(1+(k_0/2)\xi)}\right)\,d\xi^\alpha
-\frac{k_0^{1-\alpha}}{b^\alpha}\int_0^\infty\left\{
\frac{k_0^\alpha}{e^{k_0\xi}-1}
-\frac{k_0^\alpha}{k_0\xi(1+(k_0/2)\xi)}\right\}\,d\xi^\alpha.
\nonumber
\end{align}

Write
$$
c=\int_0^\infty\left(\frac1\xi-\frac1{e^\xi-1}\right) \xi^\gamma\,d\xi.
$$

After the change $k_0\xi=\eta$, we see that
\begin{align}
&\frac{ k_0^{1-\alpha}}{b^\alpha}\int_0^\infty\left\{
\frac{k_0^\alpha}{e^\eta-1}-\frac{k_0^\alpha}{\eta(1+\eta/2)}
\right\}\,d\xi^\alpha =\frac{k_0^{1-\alpha}}{b^\alpha}\int_0^\infty
\left\{\frac1{e^\eta-1}-\frac1{\eta(1+\eta/2)}\right\}d\eta^\alpha
\nonumber\\
&\qquad=
\frac{k_0^{1-\alpha}}{b^\alpha}\left\{\int_0^\infty\left(\frac1{e^\eta-1}
-\frac1\eta\right)+
\int_0^\infty\frac{d\eta^\alpha}{2(1+\frac\eta2)}\right\}=-c
\frac{k_0^{1-\alpha}}{b^\alpha}+ c_1\frac{k_0^{1-\alpha}}{b^\alpha}\,.
\end{align} 

Since
$$
\text{ $\frac1{\eta(1+\eta/2)}=\frac1\eta-\frac1{2(1+\eta/2)}$\,,}
$$
we can set
$$
c_1=\int_0^\infty\frac{d\eta^\alpha}{2(1+\frac\eta2)}\,,
$$
and write
\begin{equation}
\int_0^\infty\left(\frac1\xi -\frac1{\xi(1+\frac{k_0}2\xi)}\right)\,d\xi^\alpha
=\frac{k_0}2\int_0^\infty \frac{\,d\xi^\alpha}{1+\frac{k_0}2\xi}=
\left(\frac{k_0}2\right)^{1-\alpha}\int_0^\infty
\frac{d\eta^\alpha}{1+\eta}=c_1\left(\frac{k_0}2\right)^{1-\alpha}. 
\end{equation}
Hence,
\begin{equation}
-\frac1{b^\alpha}c_1+\frac1{b^\alpha}c\left(\frac{k_0}2\right)^{1-\alpha}
\!\!\!\!-\frac{k_0^{1-\alpha}}{b^\alpha}\int_0^\infty\!
\left\{\frac1{e^\eta-1}-\frac1{\eta(1-\frac\eta2) }\right\}d\eta^\alpha
-\frac12\int\!\frac{d\eta^\alpha}{1+\frac\eta2}\cdot
\frac{k_0^{1-\alpha}}{b^\alpha}
=-\frac1{b^\alpha}c+\frac{k_0^{1-\alpha}}{b^\alpha}c. 
\end{equation}

Since $k_0$ is the number of particles, $b=1/T$, and $\alpha=1+\gamma$, it
follows that $k_0b^\alpha$ for $\gamma>0$ is the value of the Riemann zeta
function, $\zeta(1+\gamma)$. Therefore, $k_0^{\gamma+1}$ increases for
$\gamma<1$, and the first term of the right-hand side of equation~\thetag{5}
can be neglected. Introducing the function
\begin{equation}
\mathcal M(\gamma+1)=
\bigg(\frac{c(\gamma)}{\Gamma(\gamma+1)}\bigg)^{\frac1{1+\gamma}}, 
\end{equation}
we see that the compressibility factor
$$
Z_\gamma=-\zeta(\gamma+2)/\mathcal M(\gamma+1)
$$
is subjected to a flexion\footnote{Since the ``Young moduli'' for the
compression and extension are distinct, a flexion of the spinodal occurs. If
the pressure is reduced and the temperature is not reduced, then the liquid
``begins to boil.'' The simultaneous reduction of the pressure and the
temperature makes it possible to approach the new critical point but only by
especially painstaking experiments~\cite{71}.} \; from $\gamma>0$ to
$\gamma<0$.

In this case, we obtain another critical point, which fully corresponds to the
physical meaning (see~\cite{71}).

Thus, if the compressibility factor is negative, then we divide $\mathcal{E}/N$ by
$T^{\gamma+1}$ with $\gamma <0$ rather than by $T$, because
$$
\frac{\mathcal
E}N\Big|_{\mu=0}=\frac{\zeta(\gamma+2)T^{\gamma+2}_{\mathrm{red}}} {\mathcal
M(\gamma+1)T_{\mathrm{red}}}=
T^{\gamma+1}_{\mathrm{red}}\frac{\zeta(\gamma+2)}{\mathcal M(\gamma+1)},
$$
i.e., the energy evaluated for a single particle at $\mu=0$ (at the
``degeneration'' point), for $P<0$, is proportional to
$T^{\gamma+1}_{\mathrm{red}}$, i.e., to the temperature taken to a power with
an exponent less than one. For $P<0$, the compressibility factor becomes a
dimensional quantity; however, this is always considered in this very way on
curves in the $\{Z,P\}$ space when using the van der Waals
normalization~\thetag{57}. Let us now present a graph for negative pressure for
the Lennard--Jones potential, where the new critical point is obtained by using
a computer experiment.\footnote{The absolute zero of temperature is
inaccessible. This is visually seen in the logarithmic scale of temperatures
$\ln T_{\mathrm{red}}$, where the absolute zero corresponds to~$-\infty$.}

\begin{figure}[h]
\begin{center}
\includegraphics{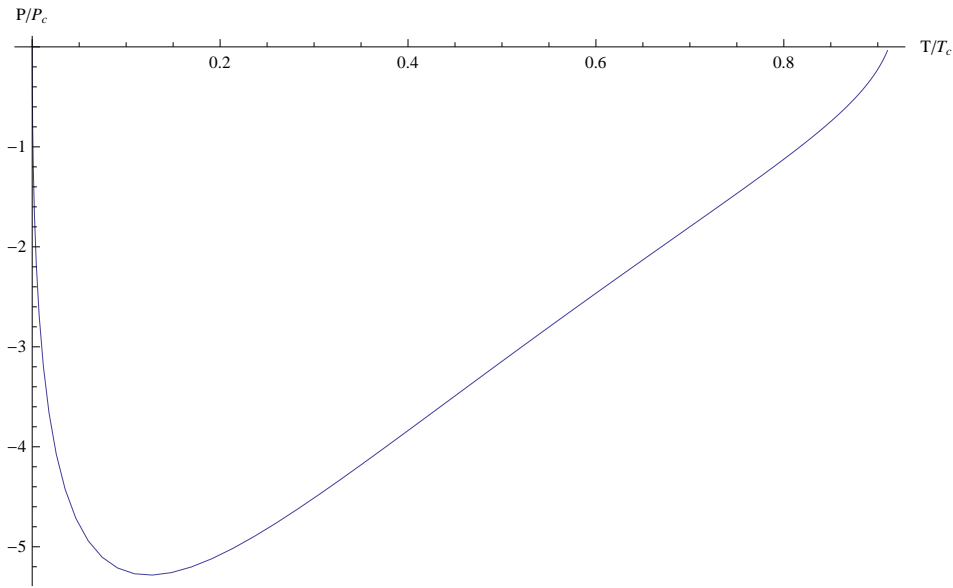}
\end{center}
\caption{The spinodal in the coordinates given by the
temperature $T_{\mathrm{red}}$ and the negative pressure $P_{\mathrm{red}}$.}
\end{figure}

\begin{figure}[h]
\begin{center}
\includegraphics{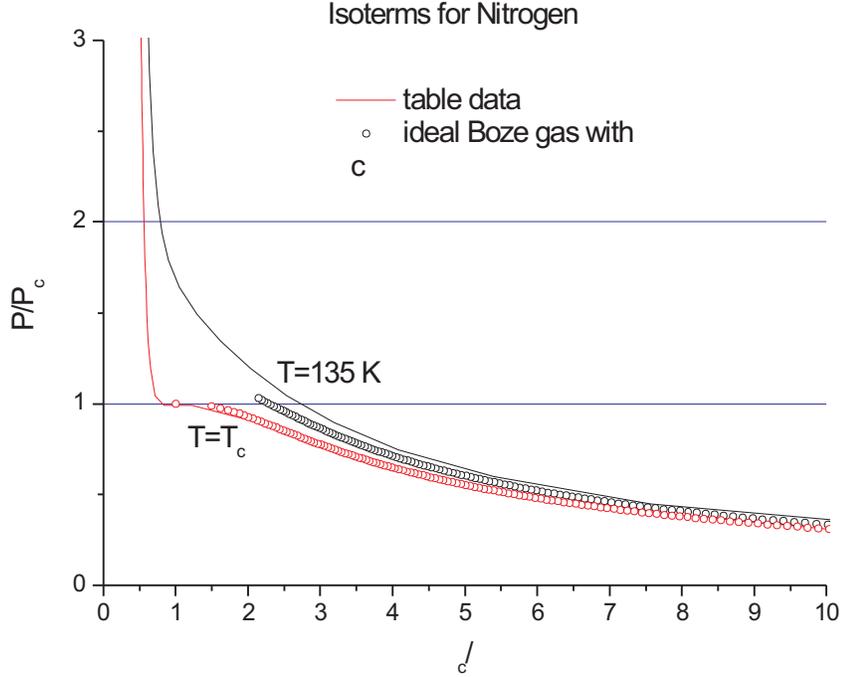}
\end{center}
\caption{Isotherms for nitrogen, $P/P_c$ versus $\rho_c/\rho$, ideal Bose gas with $\gamma=0.218$, $Zc=0.287$.}
\end{figure}

\begin{figure}[h]
\begin{center}
\includegraphics{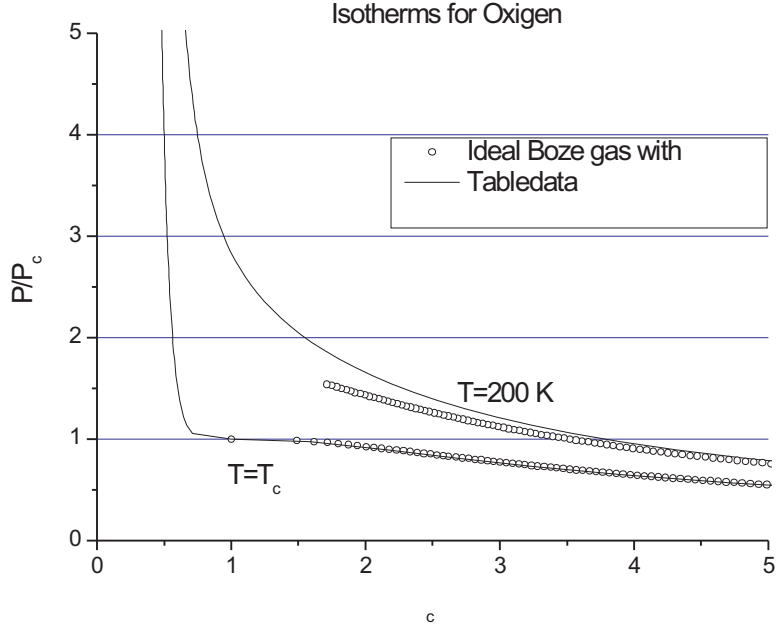}
\end{center}
\caption{Isotherms for oxygen, $P/P_c$ versus $\rho_c/\rho$, ideal Bose gas with $\gamma=0.219$.}
\end{figure}

\section{On Homogeneous Mixtures of Gases}

When considering a gas mixture, we would like to attract attention at the
following fundamental point. As is well known, in statistical physics and
thermodynamics, the energy is sometimes connected with the number of degrees of
freedom and the temperature; for example, this is the case in the
equidistribution law. It turns out in this case that the energy depends on the
temperature and on the number of degrees of freedom and does not depend on the
mass. The sequential usage of this conception gave us a continuous parameter
$\gamma$ related to the fractal dimension in the momentum space. Continuing the
use of this conception in the case of a mixture of pure gases, we are to speak
of the concentration rather than on the density, i.e., we are to neglect the
masses of miscible pure gases.

The sequential application of number theory in thermodynamics, i.e., the
consideration of the main axiom for the gas mixture (which was in fact made by
experimenters, at least in the case of air (see~\cite{71})), leads to the
formulas presented below.

If the values of energies $\mathcal E_1$ and $\mathcal E_2$ expanded into sums
of $N_1$ and $N_2$ summands, respectively, correspond to fractional
dimensions~$\gamma_1$ and~$\gamma_2$, respectively, and the values of the pairs
$\{\mathcal E_1,N_1\}$ and $\{\mathcal E_2,N_2\}$ are at the ``verge of
degeneration,'' i.e., adding an excessive number to $N_1$ and to $N_2$ leads to
the ``appearance of the Bose condensate,'' then, for the sum
 $\mathcal E_1+\mathcal E_2$ and for $N_1+N_2$, any adding an excessive number
to $N_1+N_2$ also leads to the ``appearance of the Bose condensate.''

Let $\rho_1^{\mathrm{cr}}$ and $\rho_2^{\mathrm{cr}}$ be the critical
concentrations (in the units $\mathrm{cm}^{-3}$), and let $N_1$ and $N_2$ be
proportional to the molar concentrations,
\begin{equation}
\frac{N_1}{N_1+N_2}=\alpha, 
\end{equation}
\begin{equation}
\frac{N_2}{N_1+N_2}=\beta, 
\end{equation}
\begin{equation}
N=N_1+N_2, \qquad \alpha+\beta=1. 
\end{equation}

Since
$$
\mathcal E_{\mathrm{cr}}=\mathcal E|_{\mu=0}=
N|_{\mu=0}Z^{\mathrm{cr}}\frac{\Gamma(\gamma+1)}{\Gamma(\gamma+2)}T^{\mathrm{cr}}=
N^{\mathrm{cr}}Z^{\mathrm{cr}}(\gamma^{\mathrm{cr}}+1)
$$
for $\gamma>0$, we see that, dividing the equation
$$
\mathcal E^{\mathrm{sum}}_{\mu=0}=\mathcal E^{(1)}_{\mu=0}+\mathcal
E^{(2)}_{\mu=0}=
N_1^{\mathrm{cr}}(\gamma_1+1)Z_1^{\mathrm{cr}}
T_1^{\mathrm{cr}}+N_2^{\mathrm{cr}}(\gamma_2+1)Z_2^{\mathrm{cr}}T_2^{\mathrm{cr}}
$$
by $N_{\text{sum}}^{\text{cr}}$, we obtain the relation\footnote{The total
energy does not depend on the masses of the particles, as well as in the theory
of Brownian particles (see~\cite{33}). The theoretical value is expressed in
terms of the Wiener path integral, which is equal to a Feynman path integral
with an imaginary Planck constant. This is an additional argument in favor of
the Wiener quantization.}
\begin{equation}
(\gamma_{\mathrm{sum}}^{\mathrm{cr}}+1)
Z_{\mathrm{sum}}^{\mathrm{cr}}T_{\mathrm{sum}}^{\mathrm{cr}}=
\alpha(\gamma^{\mathrm{cr}}_1+1)Z^{\mathrm{cr}}_1T^{\mathrm{cr}}_1+
\beta(\gamma^{\mathrm{cr}}_2+1)Z^{\mathrm{cr}}_2T^{\mathrm{cr}}_2, 
\end{equation}
where $Z^{\text{cr}}=\zeta(\gamma+2)/\zeta(\gamma+1)$, the subscripts 1 and 2
denote the first and the second gas of the mixture, respectively, and the
subscript $\text{sum}$ denotes the gas mixture. Similarly,
$$
S=N(Z-\mu/T),
$$
since the entropy is additive, when dividing by $N^{\text{sum}}_{\mu=0}$, we
obtain the relation
\begin{equation}
Z_{\mathrm{sum}}^{\mathrm{cr}}(\gamma_{\mathrm{sum}}^{\mathrm{cr}}+2)=
\alpha(\gamma^{\mathrm{cr}}_1+2)Z^{\mathrm{cr}}_1+
\beta(\gamma^{\mathrm{cr}}_2+2)Z^{\mathrm{cr}}_2. 
\end{equation}

It follows from the above two relations that the quantity
$\gamma=\gamma_{\mathrm{sum}}$ almost linearly depends also on the values
$\alpha$ and $T_{\mathrm{cr}}=T_{\mathrm{cr}}^{\mathrm{sum}}$. This fact is
well known as ``Kay's rule'' in the phenomenological theory of mixtures. For
air, we have $T_{\mathrm{cr}}^{\mathrm{sum}}=232\, \text{K}$, whereas
$T_{\mathrm{cr}}=255\, \text{K}$ for oxygen (20\% in air) and
$T_{\mathrm{cr}}=226\, \text{K}$ for nitrogen (80\% in air). The value of
$T_{\mathrm{cr}}^{\mathrm{sum}}$ coincides with the value of this quantity
evaluated according to the above formulas up to the accuracy of $0.5\%$.

We have defined $Z_{\mathrm{cr}}$ for the ideal gases, and hence also
$\gamma_{\mathrm{cr}}$ for a gas mixture. For a mixture of real gases, we must
define the function $\varphi_\gamma^{\mathrm{mix}}$.

It turns out that, for a mixture, the Zeno-line is not a segment of a straight
line, which is the case and is observed experimentally for pure gases.
Therefore, for a mixture, it is insufficient to find the values
$T_B^{\mathrm{mix}}$ and $\rho_B^{\mathrm{mix}}$. One must also define the
function $\varphi_{\gamma^{\mathrm{cr}}}^{\mathrm{mix}}$. This can be carried
out by using the following formulas.

We are interested only in the values of $\mu_1$ and $\mu_2$ that correspond to
the Zeno line of each of the gases (see~\thetag{52}-\thetag{53}).

Since, by assumption, the critical point enters the homogeneity domain, it
follows that the concentrations $\alpha$ and $\beta$ are preserved, and hence,
using equations~\thetag{75}--\thetag{77}, we obtain the following equation for
the sum of the entropies, where $\kappa=\mu/T$:
\begin{equation}
(\gamma_{\mathrm{sum}}+2)Z_{\gamma_{\mathrm{sum}}+2}(e^{\kappa_{\mathrm{sum}}})-
\kappa_{\mathrm{sum}}= \alpha\big\{(\gamma_1+2)Z_{\gamma_1+2}(e^{\kappa_1})-
\kappa_1\big\}+ \beta\big\{(\gamma_2+2)Z_{\gamma_2+2}(e^{\kappa_2})-
\kappa_2\big\}, 
\end{equation}
where $Z_{\gamma+2}$ is equal to the ratio
$\operatorname{Li}_{\gamma+2}(e^\kappa)/\operatorname{Li}_{\gamma+1}(e^\kappa)$.

Recall that the values of $\kappa_1$ and $\kappa_2$ are taken according to the
Zeno lines of the first and the second gas, respectively. Hence, using the
given values of $\gamma$, $\gamma_1$, and $\gamma_2$ obtained from \thetag{79},
we find~the value $\kappa=\kappa_{\mathrm{sum}}$, which defines the function
$\varphi_\gamma(V)$ by relation~\thetag{53}. This enables us to define the
dependence of $T_{\mathrm{cr}}^{\mathrm{sum}}$ for a mixture of real gases.

\begin{remark} We have proved that the Boltzmann distribution, which is
based on the fact that transpositions of two particles in a set of $N$
particles lead to a new microscopic state, is based in turn on the dogma of the
natural series, and hence cannot be applied to the concept of ``pile'' as a
macroscopic state. This implies that the Maxwell distribution describes the
ideal gas too roughly.

The introduction of new parameter $\gamma$ enables one to assign the mean
energy of the system with respect to the spectral density of a given molecule
with higher accuracy as compared with the similar assignment using the
equidistribution law with respect to the degrees of freedom, where only the
number of atoms of the given molecule is taken into account. As
$\gamma\to\infty$ or $\mu\to-\infty$ (the~chemical potential), we obtain the
old conception of ideal gas.

Certainly, the interaction of molecules depends on the spectrum of molecules.
However, the experiment shows that, for all molecules except for those of
water, there are two important constants, $T_B$ and $\rho_B$, and it turns out
that these two constants and the critical value of the compressibility factor
are sufficient to describe, with the corresponding accuracy, both the equation
of state of a pure gas and that of a gas mixture.

Bohr's Complementarity Principle, as applied to thermodynamical quantities,
enables one to elucidate the notion of critical exponents anew.
\end{remark}

\section{New approach to independent events in probability theory}

As was said above, we consider probability theory without preliminarily passing
to the limit
$$
p_i=\lim_{N\to\infty}\frac{N_i}N;
$$
we first consider the system for a finite~$N$,
\begin{equation}
M=\sum_{i=0}^\infty N_i\lambda_i, \qquad \sum_{i=0}^\infty N_i=N. 
\end{equation}
In this system of equations, we pass to the asymptotic behavior as $N\to\infty$
and $M\to\infty$ in such a way that $N$ and $M$ turn out to be connected by a
relation corresponding to relation~\thetag{14} of the present paper. We
consider the set of solutions of system~\thetag{81} as the family of elementary
equiprobable events. To be more precise, we consider the solutions of the
system of inequalities
\begin{equation}
\sum_{i=0}^\infty N_i\lambda_i\le M, \qquad \sum_{i=0}^\infty N_i=N 
\end{equation}
rather than the system of equations. As is clear from system~\thetag{13}, there
is an $N_c$ such that the number of solutions of this system is the largest
possible, and this number can asymptotically be calculated. The number $M_c$
corresponds to~$N_c$.

One can introduce two Lagrange conjugate quantities, namely, the chemical
potential $\mu$ and the inverse temperature $\beta$. The value $\mu=0$
corresponds to the critical values $N_c$ and $M_c$. Thus, we can transfer all
our considerations concerning ideal gases to the new probability theory.
Poincar\'e writes about an insufficiently clear (``obscure'') ``instinct which
we call common sense'' on which probability theory (and, I shall add, the
notion of independent events) must be based.

If one starts from the notion of ideal gas as a gas of particles without
interaction, then it is natural to assume that independent events are events
without interaction. However, when we consider a mixture of ideal gases, then
we start from the very absence of interaction between the particles. Since
relations~\thetag{78, 79} express this condition, we must use these very
relations.

In this definition of independent events, there is an {\it a priori\/} given
probability
\begin{equation}
\alpha=\frac{N_1}{N_1+N_2}, \qquad \beta=1-\alpha. 
\end{equation}
The parameter $\gamma$ is related to the notion of fractal dimension, and thus
to the Hausdorff--Besicovitch
dimension, and, in this relation, this notion, which has recently been widely
used, must necessarily be related to the new probability theory. By the
definition in~\thetag{83}, the formulas are of ``conditional'' nature in the
sense appropriate in the ordinary probability theory.

The notion of independence has already been considered in Kolmogorov's
complexity theory and completely agrees with formula~\thetag{79}. As far as
formula~\thetag{80} is concerned, it distinguishes a sufficiently important
class in the case of dependent events.

To generalize the new definition of independence of events in the new
probability theory, it is necessary to pass first of all from equiprobable
elementary events presented above to equipping these events by weights, which
is brilliantly done in Vershik's paper~\cite{12}, and to a generalization of
the new notion of independent events to Vershik's multiplicative measures. In
this case, measure theory, which is the most important element of Kolmogorov's
probability theory, will find an application also to the above new conception
of independent events.

\section{Application to economics. The theory of crisis of debts}

As already stated on numerous occasions, money obeys the ``Bose statistics'' and
banknotes of one denomination, are practically indistinguishable despite their
different numbers~\cite{3, 81}. Therefore, we can associate the ``number'' (amount)
of money with the number of particles in physics (which is consistent with
Irving Fisher's Correspondence Principle)~\cite{72}.
In economics, the goods are also averaged.
The term ``wholesale'' is usually used for them,
and the term ``at retail'' can be used only occasionally,
for example, for authorized artwork.
Fisher's Correspondence
Principle consists in the following comparison of economic and physical
quantities:

the volume of goods $Q$ corresponds to the volume of gas $V$;

the price of the amount of goods $P$ corresponds to the pressure $P$;

the amount of money $M$ corresponds to the number of particles $N$;

the turnover rate $V$ corresponds to the temperature $T$.

By Friedman's rule, the optimal amount of money corresponds to the zero nominal
percentage~\cite{73}. Combining this with the appropriately modified method of
Lagrange indeterminate multipliers, we can regard the nominal percentage~$R$ as
an addition to the amount of money~$M$.

In the same way, the value of the turnover rate $V$ is associated with the
degree of uncertainty, or entropy~\thetag{55}, which is an additional quantity:
\begin{equation}\label{p-1}
S=\frac{\pi^{\gamma+1}V^{\gamma+1}}{\Lambda^{2(^\gamma+1)}}
\bigg\{(2+\gamma)\operatorname{Li}_{2+\gamma}(z) - \operatorname{Li}_{1+\gamma}(z)\frac{R}{V}\bigg\},
\end{equation}
where $\Lambda$ is a certain constant, its own for each currency,
and $z=e^{R/V}$.

The price of the volume $Q_\gamma$ has the form
\begin{equation}\label{p-1a}
P=\frac{\pi^{\gamma+1}}{\Lambda^{2(^\gamma+1)}}V^{\gamma+2} \operatorname{Li}_{2+\gamma}(z).
\end{equation}

Now Irving Fisher's Correspondence Principle can be extended to all the
thermodynamic quantities. The ratio
\begin{equation}
\frac{PQ}{MV}\bigg|_{R=0}=\frac{\zeta(\gamma+2)}{\zeta(\gamma+1)} 
\end{equation}
allows us to obtain, for economics, the family of distributions depending on
the parameter~$\gamma$ once the value of the quantity $({PQ})/{MV}$ is
calculated as $R\to0$.

Irving Fisher had in mind the correspondence which is known today under the
title of the ``fundamental law of economics,''
$$
PQ=MV.
$$
In view of~\thetag{84}, this correspondence takes the form
\begin{equation}
PQ=Z_{\mathrm c}MV, 
\end{equation}
where the compressibility factor
$$
Z_{\mathrm c}=\frac{PQ}{MV}\bigg|_{R=0}
$$
has its own particular value for each country or region. The introduction of a
common currency for a group of countries (or regions) corresponds to the mixing
of pure gases (see~\cite{36}).

An example of the case of the zero nominal percentage~$R$ in the USSR was the
post-war reconstruction of industry accompanied by the simultaneous reduction
in prices. We see that this corresponds to the maximal entropy as a measure of
uncertainty (under a planified socialist economy!). Hence, as the turnover
rate~$V$ (similar to temperature in thermodynamics) falls, this must lead to
the equilibrium of the system of sharp stratification of society into the rich
and the poor in view of  the desire of the society to have a rest from stresses.
\footnote{This fact is one of the elements of the concept of Human
Thermodynamics.}.

Now let us introduce a general notion which is the addition of mobility braking
factors to the rate of economic processes. This may be bureaucratic red tape,
holidays, delays in taking decisions, criminal behavior, economic and natural
disasters, strikes, lack of stimulus among producers, traffic jams, flight
delays, and so on. Situations in which there are delays in the reaction of
economic agents owing to abrupt changes in policy also play a significant role
in economics.

Let us call this turnover rate braking parameter {\it viscosity} and denote
it by the letter~$\varepsilon$.

In thermodynamics, viscosity weakly depends on temperature. In economics,
viscosity depends on the degree of uncertainty (entropy). We have already
stated that viscosity for the rich can be decreased by bribes, which further
increases the stratification of society. But since, as $\varepsilon\to0$, we
obtain a sufficiently handy tool for determining various financial changes in
time (such as in tropical mathematics, which can be applied to economics), then
we can assume that this parameter is small. The widespread use of computers,
undoubtedly, decreases it still further. Since more and more computers are used
to perform various functions, which leads to a decrease of the
parameter~$\varepsilon$, it follows that the effects of the Wiener quantization
of economics corresponding to phase transitions will play an increasingly
greater role.

Therefore, we can extend Bohr's Complementarity Principle in economics by the
Wiener quantization of equilibrium economics and by geometric quantization
using the tunnel canonical operator.

 We shall consider the Wiener quantization of economics
by analogy with thermodynamics (see Section~2) using the correspondence
\begin{equation}
Q\leftrightarrow V, \qquad R\leftrightarrow\mu, \qquad M\leftrightarrow N. 
\end{equation}

If economic optimization problems (even in implicit form) involve a Hamiltonian
of the form $H(P,Q,V,S,RM)$ (see Pospelov's paper~\cite{76}), then, by
expressing $RM$ (an analog of the Gibbs thermodynamic potential) from the
condition $H=\mathrm{const}$ in terms of $P,Q,V,S$, we obtain an analog of the
equation of state.

The Wiener quantization (cf.~\cite{74}) involves the correspondence
\begin{equation}
Q=\varepsilon\frac\partial{\partial P}, \qquad
-S=\varepsilon\frac\partial{\partial V}, \qquad
M=\varepsilon\frac\partial{\partial R}. 
\end{equation}
We can assume without detriment to the asymptotics as $\varepsilon\to0$ that
the differential operators act first and the operators of multiplication by
$P,V,R$ second. More complicated equations of economics, just as the Vlasov
equations, preserve the Lagrangian property of surfaces of the type of the
``equation of state.''  Therefore, we can apply the geometric Wiener (tunnel)
quantization~\cite{22}.

The phase transition to the two-phase system ``gas--liquid'' (or ``the
poor--the rich'') occurs for a given operator rate~$V$ for a definite value of
the nominal percentage~$R$.

Debts correspond to negative values of~$M$ and~$P$. This does not preclude a
country with a huge internal debt, such as the United States at the time of the
war in Iraq, to calculate $Z_{\mathrm c}$ by formula~\thetag{84}. Then the
critical value of debt will be the optimal value of~$M$, i.e., the condition
for a maximum of the entropy for a given~$\gamma$. In the case where $P$ is
negative (as was the case in the United States when the war expenditures
produced debts) and $M$  is positive, the critical value of debt is the
continuation of $Z_{\mathrm c}$ to the negative domains of~$P$ and~$Z$. This
corresponds to negative pressure described in~\cite[Sec.~6]{77} (see also~\cite{78}).

In this case, we can also calculate the value of~$M_{\mathrm c}$ at which the
``Bose-condensate'' phenomenon, corresponding to bankruptcy, occurs
(see~\cite{79, 80}), and we can define the critical value of a debt crisis.

To calculate the parameter $\gamma$ for each currency
is an extremely difficult problem.
Since the curve of critical temperature in Fig.~10
is almost the same for a great number of gases,
the same must also hold in economics for the currencies of different countries.
This principle of correspondence between financial mathematics
and new thermodynamics of gases has already given correct forecasts.

\section*{Acknowledgments}

 The author expresses his deep gratitude for extremely useful consultations to
Russian virtuosi of experiments in physics, to V.~G.~Baidakov, V.~V.~Brazhkin,
A.~A.~Vasserman, A.~E.~Gekhman, D.~Yu.~Ivanov, V.~I.~Nedostup, and
K.~I.~Shmulovich, and also to greatest experts in applied thermodynamics, to
V.~A.~Istomin and~V.~S.~Vorob'ev, and to greatest experts in molecular physics,
to I.~V.~Melikhov, V.~N.~Ryzhov, and~A.~R.~Khokhlov. Permanent conversations
with A.~F.~Andreev, A.~V.~Chaplik, A.~I.~Osipov, S.~I.~Adyan, and
G.~L.~Litvinov helped to simplify the text and improve the style. The initial
version of the paper was studied with great attention by the late E.~G.~Maksimov.
His remarks concerning the language understandable by physicists were
especially valuable. The author also thanks D.~S.~Minenkov for the help in the
construction of the graph~9 and R.~V.~Nekrasov for the help in the construction
of the graph~12.

\end{document}